\documentclass[11pt]{article}

\usepackage{graphicx}

\usepackage{multicol} 

\usepackage{color}
\usepackage{latexsym}

\definecolor{rosso}{cmyk}{0,1,1,0.4}
\definecolor{rossos}{cmyk}{0,1,1,0.55}
\definecolor{rossoc}{cmyk}{0,0.5,1,0.2}
\definecolor{blu}{cmyk}{1,1,0,0.3}
\definecolor{blus}{cmyk}{1,1,0,0.6}
\definecolor{blucc}{cmyk}{1,0.4,0.2,0}
\definecolor{viola}{cmyk}{0,1,0,0.6}
\definecolor{viola2}{cmyk}{0,1,0.2,0.6}
\definecolor{verde}{cmyk}{0.92,0,0.59,0.25}
\definecolor{verdec}{cmyk}{0.92,0,0.59,0.15}
\definecolor{verdes}{cmyk}{0.92,0,0.59,0.4}
\font\tenrsfs=rsfs10 at 12pt
\font\sevenrsfs=rsfs7
\font\fiversfs=rsfs5
\newfam\rsfsfam
 
\textfont\rsfsfam=\tenrsfs
\scriptfont\rsfsfam=\sevenrsfs
\scriptscriptfont\rsfsfam=\fiversfs
\def\mathscr#1{{\fam\rsfsfam\relax#1}}

\def\circa#1{\,\raise.3ex\hbox{$#1$\kern-.75em\lower1ex\hbox{$\sim$}}\,}

\usepackage{amsmath,latexsym,amssymb,color,hyperref,graphicx}

\newcommand{\be}{\begin{equation}}
\newcommand{\ee}{\end{equation}}
\newcommand{\bea}{\begin{eqnarray}}
\newcommand{\ena}{\end{eqnarray}}

\renewcommand\a{\alpha}

\renewcommand\L{\ensuremath{\Lambda}}

\newcommand{\de}{\partial}

\newcommand{\ba}{\begin{eqnarray}}
\newcommand{\ea}{\end{eqnarray}}
\newcommand{\plm}{M_{\text{Pl}}} 
 
\makeatletter
\def\ps@mine{%
    \def\@oddfoot{\hfil\thepage\hfil}\let\@evenfoot\@oddfoot
    \let\@oddhead\@evenhead%
    \let\@mkboth\@gobbletwo
    \let\sectionmark\@gobble
    \let\subsectionmark\@gobble
    }
\renewcommand\section{\@startsection {section}{1}{\z@}%
                                   {-3.5ex \@plus -1ex \@minus -.2ex}%
                                   {2ex \@plus.2ex}%
                                   {\normalfont\large\sffamily\bfseries}}
\renewcommand\subsection{\@startsection {subsection}{1}{\z@}%
                                   {-3.5ex \@plus -1ex \@minus -.2ex}%
                                   {2ex \@plus.2ex}%
                                   {\normalfont\sffamily\bfseries}}
\makeatother

\numberwithin{equation}{section}

\begin{document}

\thispagestyle{empty}
\vspace*{-2.5cm}
\begin{minipage}{.45\linewidth}

\end{minipage}
\vspace{2.5cm}

\begin{center}
  {\huge\sffamily\bfseries Adiabatic Media Inflation
  }

  \end{center}
 
 \vspace{0.5cm}
 
 \begin{center} 
 {\sffamily\bfseries \large  Marco Celoria}$^{b}$,
 {\sffamily\bfseries \large  Denis Comelli}$^{c}$,
 {\sffamily\bfseries \large  Luigi Pilo}$^{d,e}$,
  {\sffamily\bfseries \large Rocco Rollo}$^{a,d}$\\[2ex]
$^a$ Gran Sasso Science Institute (GSSI)\\Viale Francesco Crispi 7, I-67100 L'Aquila, Italy\\\vspace{0.3cm}
$^b$ ICTP, International Centre for Theoretical Physics Strada Costiera 11, 34151, Trieste, Italy \\\vspace{0.3cm}
$^c$ INFN, Sezione di Ferrara, I-44122 Ferrara, Italy \\\vspace{0.3cm}
$^d$INFN, Laboratori Nazionali del Gran Sasso, I-67010 Assergi, Italy\\\vspace{0.3cm}
$^e$Dipartimento di Ingegneria e Scienze dell'Informazione e Matematica, Universit\`a degli Studi dell'Aquila,  I-67010 L'Aquila, Italy\\\vspace{0.3cm}
 {\tt mceloria@ictp.it}, {\tt comelli@fe.infn.it}, {\tt luigi.pilo@aquila.infn.it}, {\tt rocco.rollo@gssi.it},
\end{center}

\vspace{0.7cm}

\begin{center}
{\small \today}
\end{center}

\vspace{0.7cm}

\abstract{ We study the dynamics of inflation driven by an adiabatic
  self-gravitating medium, extending the previous works on fluid and
  solid inflation.  Such a class of media  comprises  perfect fluids, zero
  and finite temperature solids. By using an effective field 
  theory description, we compute the power spectrum for
  the scalar curvature perturbation   of constant energy
  density hypersurface $\zeta$ and the comoving
  scalar curvature perturbation ${\cal R}$  in the case of slow-roll, super
  slow-roll and  $w$-media inflation, an inflationary phase with $w$
  constant    in the range   $-1 <w <-1/3$. A similar computation is done for the
  tensor modes. Adiabatic media are characterized by intrinsic entropy
  perturbations  that can give a significant contribution to the power
  spectrum and can be used to generate the required seed  for  primordial
  black holes.
 For such a media, the Weinberg theorem is typically violated
  and 
 on super horizon scales neither $\zeta$ nor ${\cal R}$ are
  conserved and moreover $\zeta \neq {\cal R}$. 
 Reheating  becomes
  crucial to predict the spectrum of the imprinted primordial
  perturbations. We study how the difference between $\zeta$ and
  ${\cal R}$ during inflation gives rise to relative entropic
  perturbations  in $\Lambda$CDM.
 }
\clearpage

\section{Introduction}
Inflation is probably the most successful way to solve the horizon
problem in cosmology and, at the same time, to give a simple explanation of
the indirectly observed spectrum of primordial perturbations. In the
case of single field,  inflationary
predictions are largely  independent on the details of
  the  reheating phase. Indeed, according to   the  Weinberg
theorem~\cite{Weinberg:2003sw,Weinberg:2008zzc}, irrespective of the
constituents of the Universe, there is always an adiabatic mode,
constant on superhorizon scales, which can be identified with the
comoving curvature perturbation ${\cal R}$, or equivalent with the
uniform curvature perturbation $\zeta$. When the theorem holds, 
${\cal R}$ or equivalently $\zeta$ can be used to set the initial conditions 
for the perturbations in the thermal radiation dominated phase of the Universe. 
Thus, the power of the Weinberg theorem relies in the possibility of
simply inflate and forget about the largely  unknown period when
  the  inflaton
dominated Universe is turned into a radiation dominated one.
However, going beyond single field inflation can trigger violation of the Weinberg
theorem; for instance,  multi-field inflation will typically produce a
mixture of adiabatic and isocurvature primordial perturbations, as a
result ${\cal R}$ and $\zeta$ will not be constant on super-horizon scales.
 In this framework,  it is very useful 
 to  explore the general physical implications on symmetry grounds
instead of dealing with a plethora of different models.  In the case
of single clock inflation, the inflaton itself can be used as a ``clock''
 that breaks  spontaneously time reparametrization invariance of general
relativity (GR); the  effective action description was given in~\cite{Cheung:2007st},
see also~\cite{Weinberg:2008hq} for a different formulation.

More subtle violations can take place when   neither
  ${\cal R}$ nor $\zeta$
are  conserved at superhorizon scales \cite{Kinney:2005vj,Namjoo:2012aa,Motohashi:2014ppa,Akhshik:2015nfa,Celoria:2017xos}. This
is the case, for instance, for fluid~\cite{Chen:2013kta} and solid
inflation~\cite{Endlich:2012pz}. In the present work, we present a general
set up capable to give an effective field theory description of
inflation driven by a generic adiabatic, non-dissipative medium.
Such effective theory
description~\cite{Matarrese:1984zw,Dubovsky:2005xd,Dubovsky:2011sj,Ballesteros:2012kv,Ballesteros:2016kdx,Celoria:2017bbh}
is based   on   four scalar fields $\{\varphi^A
\, , \;  A=0,1,2,3 \}$,  and comprise fluid inflation and solid
inflation and their generalization, namely non-barotropic fluid
inflation  and finite temperature solid inflation. The basic
assumption is that the entropy per particle
perturbation $\delta \sigma$, when present, is conserved in time.
 It is worth to point out that, though $\delta \sigma$
is time independent, it can give an important contribution to the power
spectrum; in particular, it can be used as initial seed of perturbation for the
generation of primordial black holes that could constitute a sizable
amount of dark matter in our universe~\cite{Bird:2016dcv,Sasaki:2016jop, Clesse:2016ajp,Espinosa:2017sgp}.
The case of a constant in time $\delta \sigma$ also constitutes
a simplified dynamical setup to study a generic
supersolid~\footnote{The detailed analysis will be given in a
separate paper~\cite{future}.}, the most general case of a
 non-dissipative  
 self-gravitating medium, which 
has two propagating scalar modes and one is precisely $\delta
\sigma$~\cite{Celoria:2017bbh}. Indeed, in this case the entropy
perturbations becomes constant at superhorizon scales in a wide range
of  parameters.

The outline of the paper is the following.
In section \ref{adiabatic_media} we briefly review the effective field
theory description of adiabatic  self-gravitating media.
In section \ref{Toy_model}, the dynamics of ${\cal R}$ and $\zeta$ is
derived in general terms, discussing possible sources of violation of the
Weinberg theorem. Section 
\ref{GenSol} is devoted to study the dynamics of inflation driven by 
an adiabatic medium giving the general expression of ${\cal R}$ and
$\zeta$ and the corresponding power spectrum; in particular the role of entropic perturbation is studied.
In section \ref{Media-inflation} we give a detailed analysis of
inflation for barotropic and non-barotropic fluids, zero-temperature
and finite temperature solids in the following regimes of interest:
slow-roll, super slow-roll and $w$-media inflation.
Section \ref{G_W} deals with tensor modes during inflation and the
resulting  tensor-scalar ratio.
Section \ref{Post-Inf} is devoted to the study of the post inflationary
phase that leads to a radiation dominated Universe by computing the final values
of ${\cal R}$ and $\zeta$.
Our conclusions are given in Section \ref{Concl}.  

\section{Self-Gravitating  Media}
\label{adiabatic_media}

The effective field theory (EFT) description of the dynamics of a generic
non-dissipative medium can be used to systematically describes the
symmetry breaking pattern of spacetime symmetry during inflation.  The
EFT is based on four Stuckelberg scalar fields $\varphi^A$, $
A=0\,,1\,,\,2\,,\,3$, related to the Goldstone bosons for
the spontaneous breaking of spacetime
translations~\cite{Dubovsky:2005xd,Celoria:2017bbh}.  The action is
required to be invariant under shift symmetry  $\varphi^A \rightarrow \varphi^A +
c^A$. Among the set of scalar operators containing a single derivative of $\varphi^A$
\be
\{ C^{AB} = g^{\mu \nu} \de_\mu \varphi^A \, \de_\nu \varphi^B \;\qquad
 A,B=0,1,2,3 \} \, ,
\ee
imposing internal $SO(3)$ invariance selects 8 independent
operators~\footnote{ Actually $\pmb{B}^3$ can expressed in terms of
  $\pmb{B}^2$, $\pmb{B}$ and the identity $\pmb{I}$ by using the Caley-Hamilton theorem; thus
  in this sense $y_3$ is redundant.}  
that can enter in the EFT action~\cite{Celoria:2017bbh,ussgf}
\be
\begin{split}
& X=C^{00}   , \quad  Y =u^\mu \de_\mu \varphi^0, \quad \tau_i=
\text{Tr} \left[\textbf{B}^i\right]  \; \;  i=1,2,3  , \quad  y_n =\text{Tr} \left(\pmb{B}^n
  \cdot \pmb{Z} \right)  \; \;  n=0,1,2,3 ,\\
&  \left(\textbf{B}\right)^{ab}=C^{ab}  , \qquad Z^{ab}= C^{a0} \,
C^{b0} \quad \; a=1,2,3  
\, ;
\end{split}
\ee
where we have introduced the four velocity of the medium $u^\mu$
 such
that $u^\mu\partial_\mu\varphi^a=0$ 
\be
u^\mu=
-\frac{\epsilon^{\mu\nu\alpha\beta}}{6\;b\;\sqrt{-g}}\,\epsilon_{abc}\;\partial_\mu\varphi^a\,\partial_\nu\varphi^b\,\partial_\beta\varphi^c
\, , \qquad u^2=-1 \, .
\ee
The $b$ operator is given by the following
combination of $\tau_i$
\be
b =  \text{Det}(\pmb{B})^{1/2}  = \left( \frac{\tau_1^3 -3 \,  \tau_1 \, \tau_2 +2 \, 
    \tau_3}{6}\right)^{1/2} \, .
\ee
The most general non-dissipative medium minimally coupled with gravity
(self-gravitating) is described by the action
\be
S=M_{pl}^2 \int d^4 x \, \sqrt{-g} \;R+\int d^4 x \, \sqrt{-g}\,
U(X,\, Y, \, \tau_i, \, y_n)\, .
\label{act}
\ee  
whose energy density and isotropic pressure are given by
\ba\label{ed}
&&\rho= -U+Y\;U_Y-2\;Y^2\; U_X
\\
&& p=U-b\;U_b-\frac{2}{3}\;U_X\;(Y^2+X)-\frac{2}{3}\;\sum_{n=1}^3\;n\;\tau_n\;U_{\tau_n}
-\frac{2}{3}\,\sum_{n=0}^3\;(n+1)\;y_n\;U_{y_n} \, .
\label{pre}
\ea
 The entropy density $s$ can be expressed in terms of the
Noether current $J^0_\mu$ of the shift symmetry of $\varphi^0$ as
\be 
s=-u^\alpha \, J_{\alpha}^0 =U_Y-2\;Y\;U_X \, ,
\ee
and the temperature $T$ is represented by the operator $Y$.
The number density $n$ can be identified as the projection along the four-velocity of an off shell  conserved current
\be
 n=-u_\alpha \, J^\alpha\,,\qquad
J_\alpha= b \,u_\alpha\,, \qquad \nabla^\alpha \, J_\alpha=0\, .
\ee 
 Finally,    the  conservation of  entropy per particle $\sigma=s/n$ along the flow lines   gives
 \be
 \label{sig}
  u^\a\nabla_\a
\sigma=-\frac{2}{b}\;\nabla^\a\;\left[
  \;U_X \left( Y \, u_\alpha +\partial_\alpha \varphi^0\right)
  +\;\sum_{n=0}^3 \; U_{y_n}\;
   {\cal C}^{0a} \; ({B}^n)^{ab} \; \partial_\a  {\varphi}^b\right]
 \, .
 \ee
Internal symmetries can be used to enforce some special thermodynamical and
dynamical properties of the medium.
The mechanical properties are obtained analysing the structure of the
EMT (a full analysis is   given
in~\cite{Ballesteros:2016kdx,Celoria:2017bbh}).  It is useful to summarize the
main features associated with the presence/absence of some of the
operators in the Lagrangian which corresponds to specific internal symmetries.
\begin{itemize}
\item Perfect Fluids: 
\begin{itemize}
\item $U(b,\,Y) :$
only $\{\varphi^a \, , \; \; a=1,2,3 \}$  are present;  the Lagrangian is invariant  under internal volume
preserving diffeomorphisms $V_s\text{Diff}$:
$\varphi^a\to\Psi^a(\varphi^b)$\,,\quad $\det |\partial \Psi^a
/\partial{\varphi^b}|=1$, $a,b=1,2,3$.
\item $U(X):$ 
  it is the most general Lagrangian for a perfect irrotational fluid  with only $\varphi^0$    present.
\end{itemize}
\item Superfluids 
  $U(b,\,Y,\,X):$ 
  invariant under transformations of $\varphi^a$  corresponding to
  $V_s\text{Diff}$ and $\varphi^0\to\varphi^0+f(\varphi^a)$.
\item Solids   $U(Y,\,\tau_i):$
 most general Lagrangian  with only $\{\varphi^a \, , \; \; a=1,2,3 \}$ present.
\item Supersolids 
  $U(Y,\,X,\,\tau_i,\,y_n)$. 
\end{itemize}  
The main thermodynamical characterization  of the medium is based on  (\ref{sig}).
 \begin{itemize}
 \item When $u^\a\nabla_\a
\sigma=0$ the medium is named {\it adiabatic}.\\
The conservation of $\sigma$ can be enforced by a symmetry which
 forbids the presence of $X$ and $y_n$ in the Lagrangian. This is the case for the $\varphi^a$-dependent shift of $\varphi^0$, namely
\be
\varphi^0 \to \varphi^0 + f(\varphi^a) \, , \qquad a=1,2,3 \, ;
\label{sym}
\ee
where only $Y$ and $\tau_i$ are left invariant (and also the operator $b$). 
In such a way we can identify as adiabatic media: finite temperature
non-barotropoic perfect fluids described by the Lagrangian  $U(b,\,Y )$ and 
finite temperature solids described by $U( Y, \,
\tau_i)$.
\item When $\sigma=0$ the medium is named {\it isentropic}.
 \\
 Isentropic media are characterized by the absence of $Y$ and $X$ in the Lagrangian.
 Such a class of media includes perfect fluids $U(b)$, solids $U(\tau_n)$ and some specific
 superfluids and supersolids see ~\cite{Celoria:2017bbh}.
 \end{itemize}
Notice that in a FRW   background, if $\sigma$ is constant in time, it  can have a
non-trivial spatial distribution.\\
In  the rest of this paper we will
only consider adiabatic media; the general case with a dynamical
$\sigma$ will be studied elsewhere~\cite{future}.\\
In flat spacetime, the dynamics of an adiabatic medium can be found 
by studying the fluctuations around a rotational invariant
configuration by taking 
\be
g_{\mu \nu} =\eta_{\mu \nu} , \qquad \varphi^0 = t + \pi_0 \, , \qquad
\varphi^i=x^i+\partial^i \pi_L+\pi_T^i \,, \qquad \de_i  \pi_T^i=0 \,  .
\ee
At the quadratic level we have
\be
\begin{split}
{\cal S}^{(2)} &= \plm^2 \int d^4 x \Big [ \frac{(\bar{\rho}+\bar{p})}{2} \,\de_t\pi_L\, \Delta\,
  \de_t \pi_L + \left(M_3- M_2 \right) \pi_L \Delta^2 \pi_L+ M_0 \,
  {\pi_0'}^2+ 2\,  M_4 \, \pi_0' \Delta \pi_l\\
&  +\frac{(\bar{\rho} +\bar{ p})}{2}\,
  \de_t \pi_T^i \,\de_t \pi_T^i - \frac{M_2}{2} \pi_T^i\, \Delta \pi_T^i
\Big ] \, ;
\end{split}
\ee
where $\bar{\rho}$ and $\bar{p}$ are the background values of the energy
  density and pressure (constant in space and time) while $\Delta = \delta_{ij} \de_i \de_j$;  
finally  the time derivative of a
function $f$ is denoted by $f'$.  We have also introduced the
mass parameters $\{ M_A \, ; A=0,1,2,3,4\}$  which
characterize the quadratic action of a self-gravitating medium; their definition in terms of the
derivative of $U$ is given in appendix \ref{MXY}.
The main feature of the symmetry (\ref{sym}) is that
the temporal Stuckelberg field $\pi_0$ enters in the quadratic action only through its time
derivative and one can make a field redefinition by setting $\pi_0'=\Pi$,
then $\Pi$ can be integrated out from the quadratic action by using its
algebraic equation of motion; namely
\be
\Pi = c_b^2 \, \Delta \pi_L + F(\vec{x})\, \,,
\label{temps}
\ee
where $ F(\vec{x})$ is an arbitrary function of the spatial
coordinates $x^i$. 
One then gets~\footnote{For simplicity, in (\ref{quadact}) we have not taken into
  account the presence of $F$ in the solution of $\Pi$. The net effect
  is to produce in the quadratic action an irrelevant constant term plus
  a linear term which alters the solution of the equation of motion for
  $\pi_L$ making room for $\delta \sigma$; actually $F= \frac{\delta
    \sigma}{2 \, \plm^2 \, M_0}$, see (\ref{anIM}). Neither  the propagation
  velocity nor the stability are influenced by $F$.} 
\be
{\cal S}^{(2)} = \plm^2 \int d^4 x  \, (\bar{\rho}+\bar{ p})\,\Big [\frac{1}{2}\de_t\pi_L \Delta
  \de_t \pi_L - \frac{c_L^2 \,  }{2}\pi_L \Delta^2 \pi_L +\frac{1}{2}
  \de_t \pi_T^i \de_t \pi_T^i - \frac{M_2}{2\, (\bar{\rho}+\bar{ p})} \pi_T^i \Delta \pi_T^i
  \Big ] \, ;
  \label{quadact}
\ee
with 
\be
\label{cb}
c_b^2 = - \frac{M_4}{M_0} \, .
\ee
There the longitudinal modes propagate with velocity
\be
\label{cl}
c_L^2 =\frac{2 \, \plm^2 \left[M_4^2+M_0
    \left(M_2-M_3\right)\right]}{M_0 \, 
 (\bar{\rho}+\bar{ p})} \equiv c_s^2 + \frac{4}{3} \, c_T^2 \, ,
\ee
 where
 \be\label{cs}
 c_s^2 = \frac{2 \, \plm^2 \left[3 M_4^2+M_0 \left(M_2-3
       M_3\right)\right]}{3 \, M_0 \, 
   (p+\rho )}
 \ee
 is the adiabatic sound speed, while
 \be
 \label{cT}
 c_T^2 =  \frac{\plm^2 \, M_2}{(\bar{\rho}+\bar{ p})} 
 \ee
 is the propagation speed of the transverse modes.

 The case $M_0=0$~\footnote{Actually $M_0=0$ also implies $M_4=0$.} is special and corresponds to an enhanced symmetry
 which turns the adiabatic medium into an isentropic
 one~\cite{Celoria:2017bbh}. The quadratic action can be obtained by
 simply setting $\pi_0' =0$ in (\ref{quadact}); the
 longitudinal modes propagate with velocity  $c_L^2$ where the adiabatic sound speed $c_s^2$ this time is given by 
\be 
c_s^2 = \frac{2 \, \plm^2(M_2 - 3 \, M_3)}{9 \, (\bar{\rho}+\bar{ p})} \, .
\ee
Stability in flat spacetime requires that
\be
(\bar{\rho}+\bar{ p}) >0  \, \qquad M_3>M_2 \, , \qquad M_2 >0 \, ,
\ee
or equivalently $(\bar{\rho}+\bar{ p})>0$ and $c_L^2 >0$, $c_T^2 >0$ which is the
null energy condition
augmented by the positivity
relations for the mass parameters entering in the dynamics of the
fluctuations of the medium,
see~\cite{Ballesteros:2016kdx,Celoria:2017bbh}.
 Notice that when the operator $X= g^{\mu \nu} \de_\mu \varphi^0
\de_\nu \varphi^0$ is present in the Lagrangian, $\pi_0$ is a genuine
propagating mode (at the quadratic level this happens when $M_1 \neq
0$) and the medium is non longer adiabatic and it will feature two
scalars and two vectors propagating modes.

\section{\texorpdfstring{$\zeta$}{Lg} and \texorpdfstring{${\cal R}$}{Lg} Evolution}
\label{Toy_model}
The scalar part of the perturbed metric in the conformal
Newtonian gauge takes the form
\be
g_{00}=a^2(-1+2 \, \psi)\, , \qquad g_{0i} =0 \, , \qquad g_{ij}=
a^2(1+  2 \, \phi )\delta_{ij} \, ,
\ee
\be
\varphi^0 = \bar{\varphi}(t) + \pi_0 \, , \qquad \varphi^i = x^i +
\de_i \pi_L \, .
\ee
  We employ conformal time and, as before, time derivative of a
  function $f$ is denoted by $f'$.   
A generic medium will sport both density and entropy perturbations,
a general discussion can be found
in~\cite{Ballesteros:2016kdx,Celoria:2017bbh}, the 
relation with massive gravity is discussed  in~\cite{ussgf,Celoria:2017hfd}.
For our purpose, we just need to  consider perturbations
of adiabatic  media. By using the EFT for
isentropic media, one can write in a closed form the equations governing the dynamics 
of the comoving and uniform curvature gauge invariant perturbations defined by
\be
{\cal R}=- \phi  -  {\cal H} \, v  , \qquad
\qquad \zeta=-\phi -\frac{\delta \rho}{3( \bar \rho+\bar p)}\, ;
\ee
where $v$ is the scalar part of the medium's velocity,   $\bar p$ and $\bar \rho$ are the background values 
while ${\cal H} = a'/a$ is the conformal Hubble parameter.
 The dynamics of perturbation is encoded in very few
parameters, namely the equation of state $w = \bar p /\bar \rho$, the
various sound velocities defined in (\ref{cb}), (\ref{cs}),
  (\ref{cT}) and (\ref{cl}).
  In general, the  pressure perturbation can be written as
\be
\delta p = c_s^2 \, \delta \rho + \Gamma \, ;
\label{nonad}
\ee
which define the non-adiabatic contribution $\Gamma$. 
The scalar perturbations ${\cal R}$, $\zeta$,  the scalar anisotropic
stress $\Xi$ and $\Gamma$ are related by
\be
\zeta =\mathcal{R}+ \frac{a^2 \,  \Gamma}{18 \, \plm^2 \, (w+1)\, \mathcal{H}^2\,
   c_s^2}+\frac{k^2 \, \Xi }{27\, (w+1) \,\mathcal{H}^2\,
   c_s^2}-\frac{\mathcal{R}'}{3 \,\mathcal{H}\, c_s^2} \, ,
\label{Rzrel}
\ee
 By using the EFT for the medium, one can derive the expression of the
energy density and pressure perturbations $\delta \rho$ and $\delta
p$, the scalar part $\Xi$ of the anisotropic stress and the
perturbation $\delta \sigma$ of the entropy
per particle in terms of the Stuckelberg fields~\cite{Celoria:2017bbh}; namely
\bea
&& {\delta \rho}=- \frac{6\, \plm^2\,\mathcal{H}^2}{a^2}\left( 3\, \phi+k^2 \,\pi _L \right) \;
   (1+w)+\frac{\bar{\varphi}'}{a^4}\;\delta\sigma \, ; \label{deltap}\\ 
 &&\delta p =-  \frac{6\, \plm^2\,\mathcal{H}^2}{a^2}\,c_s^2\;\left( 3\, \phi+k^2 \,\pi _L \right) \;
 (1+w)+\frac{c_b^2\; \bar{\varphi}'}{a^4}\;\delta\sigma \, ;\\[.2cm]
 &&  \Xi=2 \,M_{pl}^2 \,(\phi-\psi) =-\frac{2 \,M_{pl}^2 \, M_2 }{a^2} \,\pi_L \, ; \label{anIM}\\
 && \delta \sigma = 2 \, \plm^2 \,\frac{M_0}{\bar{\varphi}'}\, \left[ \psi +
  \frac{\pi_0'}{\bar{\varphi}'} + c_b^2 \left(3 \, \phi  + k^2 \,  \pi_L
  \right) \right] \, .
\label{deltas}
\ea
  By comparison of  (\ref{deltap}), (\ref{deltas}) and (\ref{nonad}) we deduce that
\be
\Gamma=\frac{\bar{\varphi}'  \,\left( c_b^2  -  c_s^2\right)}{
  a^4}\;\delta\sigma \,.
\label{Gamma}
\ee
The dynamics of $\sigma$ is such that
\be
\delta \sigma'=\frac{k^2 \, \plm^2 \, M_1\, \left(\pi _0-\pi _L' \,\phi
    '\right)}{\phi '{}^2} \, ;
\ee
in agreement with the fact that when the medium is adiabatic, as enforced
by (\ref{sym}), then $M_1=0$ and $\sigma ' =0$.
Though $\delta \sigma$ does not depend on time, in general it will be a
nontrivial function $\delta\sigma(\vec x)$ of the spatial coordinates
$\vec{x}$ or of  the comoving momentum $\vec{k}$  in Fourier space. $\delta \sigma$ becomes a
propagating mode when the symmetry (\ref{sym}) is not enforced and the medium is not
adiabatic anymore. We stress that, even in the general case, where $\delta \sigma$ is propagating, the entropy perturbations are constant on superhorizon scales under mild assumptions, see appendix \ref{sigma_time_dep}.\\
 The uniform curvature perturbation $\zeta$ can be
  expressed in terms of $\delta \sigma$ and the longitudinal phonon mode $\pi_L$  
\be
\zeta= \frac{k^2}{3}\, \pi_L - \frac{\bar{\varphi}'}{3 \, a^4\,(\bar{\rho}+\bar{ p})}\, \delta
\sigma \, .
\label{zetapi}
\ee
  Then  by using the expressions for the
anisotropic stress (\ref{anIM}) and for $\delta \sigma$ (\ref{deltas})
and (\ref{Gamma}) given by EFT,   one gets the following relation among
uniform and comoving curvature perturbations and entropy per particle perturbation 
\be
\frac{ \bar{\varphi}'\, \left(c_L^2-c_{b
   }^2\right)}{18 \,a^2\, \plm^2 \,(w+1)\, \mathcal{H}^2\,
   c_L^2}\, \delta \sigma -\mathcal{R} \,\frac{c_s^2}{c_L^2}+\frac{\mathcal{R}'}{3 \, \mathcal{H} \, 
 c_L^2}+\zeta =0 \, .
\label{Rzetarel}
\ee
The perturbed Einstein equations lead to the following dynamical
equation for ${\cal R}$
\be
\left( {\cal R}' \,L_{\cal R} \right)' + m_{\cal R}^2 \, {\cal R} +
A_\sigma \, \delta \sigma=0
\, , \qquad  \qquad L_{\cal R}=\frac{ a^2\,
  (w+1)}{c_L^2} \, ;
\label{eqR}
\ee
where 
\be
\begin{split}
&  m_{\cal R}^2 = \frac{a^2 \, (1+w)}{c_L^4} \left \{k^2 \, c_L^4+2 \,
  \mathcal{H} \, 
  c_L^2 \left[3 \, (w+1) \, 
    \mathcal{H} \, c_T^2+2 \, c_T^2{}'\right]-4 \, \mathcal{H}\,  c_T^2 \, 
 c_L^2{}'\right \} \, ; \\
& A_\sigma = \frac{\bar{\varphi}'}{M_{pl}^2} \left [\frac{ 
      (w+1) \left(c_L^2-c_b^2\right)+2 \, c_b^2 \left(c_b^2-c_s^2\right) }{4  \, c_L^2} - \frac{ c_b^2{}'}{6 \, 
   \mathcal{H} \, c_L^2}+ \frac{c_b^2 \, c_L^2{}'}{6 \, 
   \mathcal{H} \, c_L^4} \right ]\, .
 \end{split}
\ee
As discussed in section \ref{adiabatic_media},  $c_T$, $c_L$ and $c_b$ are
functions of the mass parameters $\{ M_A \, , \; A=0,1,2,3,4 \}$ which
are expressed in terms of first and second derivatives 
 of the Lagrangian describing the medium, see appendix \ref{massesapp}
 and~\cite{Celoria:2017bbh}. A similar equation for
 $\zeta$  can be obtained by using (\ref{Rzrel}) and it is given in
 appendix \ref{app:zeta}.
 \\
 By using the field redefinition ${\cal R}_c=L_{\cal R}^{\frac{1}{2}} \;{\cal R}$, the evolution eq. (\ref{eqR}) for ${\cal R}$~\footnote{The same
  is true for the $\zeta$ evolution equation.} can be recast  in the following
canonical form 
\be
 {\cal R}_c'' + m_{{\cal R}_c}^2 \,  {\cal R}_c +
 \frac{A_\sigma}{L_{\cal R} ^{1/2}} \, \delta \sigma =0 \, , \qquad \qquad m_{ {\cal R}_c}^2 =
 \frac{m_ {\cal R}^2}{L_{\cal R} }-\frac{L_{\cal R} ''}{2 \, L_{\cal R} }+\frac{L_{\cal R} '{}^2}{4 \, L_{\cal R} ^2}
 \, .
 \label{caneq}
 \ee
 The above equation  can be used to study the
violation of the Weinberg theorem; indeed when $\delta\sigma=0$ (isentropic medium), the curvature perturbations ${\cal R}$ or $\zeta$ can grow at
superhorizon scales if
\begin{enumerate}
  \item the anisotropic stress is such that in the limit $k \to 0$,
    ${\cal R} =$constant is not a solution of (\ref{eqR});
  \item
  the background evolution is such that the would-be decreasing mode is
  actually increasing.
\end{enumerate}
Condition 1 is realized  when $M_2 \neq 0$, e.g. for a solid-like 
media, while condition 2  is realized in a
super slow-roll phase when the slow-roll parameter $\epsilon$ goes to
zero extremely fast.

\section{Inflationary Dynamics}
\label{GenSol}

Inflation~\cite{Guth:1980zm,Linde:1981mu,Albrecht:1982wi}, 
in  its slow-roll single field incarnation, 
predicts an almost Gaussian and scale-free spectrum of primordial
perturbations. An important point, specially for the study of
non-Gaussianity, is the pattern of spontaneous  breaking of spacetime
diffeomorphisms triggered by vacuum configuration of the fields
responsible of inflation. For instance, in the case of single field
inflation, time reparametrization is broken  while spatial diffs are
unbroken~\cite{Cheung:2007st}.  Our approach to the  dynamics
of self-gravitating media allows us to extend to such large class of
systems the analysis of~\cite{Cheung:2007st} for single field
inflation and~\cite{Endlich:2012pz} for solids. We stress that our EFT
description of self-gravitating media is 
rather different and in some sense complementary to multi-field
inflation, see for instance~\cite{Langlois:2008qf,Arroja:2008yy}.
Indeed, the different class of media are controlled by symmetries
which also determine the breaking patter. Moreover, the
equations of motion for  the Stuckelberg fields are equivalent to the
EMT   conservation. 
 As discussed in \ref{Toy_model}, the entropy
per particle perturbation, constant in time for adiabatic media, will
constitute an additional source, besides the anisotropic stress, to the violation of
the Weinberg theorem. 
Let us define the standard inflationary parameters
\be
\epsilon=
1- \frac{{\cal H}'}{{\cal H}^2} =\frac{3}{2}\,(1+w)\, ,
\qquad \eta=
  \frac{\epsilon'}{{\cal H} \,  \epsilon} \, ;
\label{slow_param}
\ee
inflation sets in when $w < -1/3$ or equivalently when $\epsilon
<1$. Slow-roll
is characterized by $\epsilon, \, \eta \ll 1$ but here we will
consider also regimes in which $\epsilon$ and $\eta$ do not need to be
small. In general,  $\epsilon$ and $\eta$ determine also the
adiabatic sound speed 
\be
c_s^2=w-\frac{w'}{3 \,{\cal H}\,(1+w)}= -1+\frac{2}{3} \, \epsilon -\frac{1}{3}\,\eta   \, .
\label{speed}
\ee
 The other independent parameter is $c_L^2$ and it will be taken  as a constant.\\
Adiabatic  media can be divided in two large classes: fluids and
solids. The main difference is that while for a fluid the energy of a
generic element depends only on its volume, for a solid this is not
the case and also the shape matters. 
From the point of view of
symmetries, fluids  have an infinite number of internal symmetries
corresponding to volume preserving internal diffeomorphisms. 
The net result is that
for a fluid the anisotropic stress is zero and then $M_2=0$, moreover
$c_L^2 = c_s^2$. 
For the
class of media we are considering the pattern of spontaneous breaking
of spacetime symmetries is rather different from single field inflation,
where time diffeomorphisms are broken while spatial ones
are preserved~\cite{Cheung:2007st}.
 For an isentropic medium it is just the
opposite:  time diffeomorphisms are unbroken and spatial  diffeomorphisms
are broken~\cite{Endlich:2012pz}. 
In the case of the most general
adiabatic media, both temporal and spatial diffeomorphisms are broken.
In general there is a single scalar mode that propagates;  the seed for primordial perturbations is generated by quantum
fluctuations of the canonically quantized field ${\cal R}$ or $\zeta$
in  the Bunch-Davies vacuum
that become classical after
horizon exit. The dynamical equation for ${\cal R}$  in the
canonical form  (\ref{caneq}) is such that the mass term
$m_{{\cal R}_c}^2$ gives the same propagation speed  of  $\zeta$ in the UV, namely
\be
\lim_{k \to \infty} k^{-2} m_{{\cal R}_c}^2 = c_L^2 \, .
\label{zeta-vel}
\ee
Thus, the canonical
quantization of both ${\cal R}$ and $\zeta$ singles out the
Bunch-Davies vacuum if their positive frequency mode, for large $k$, is
proportional to $\exp(- i \, c_L\, k \, t)$. We will always
make this choice in the present paper. Of course, as minimal requirement,
we should have
\be
c_L^2 > 0 \, , 
\label{pos}
\ee
and to avoid superluminality $c_L^2<1$. 
For each class of media we will analyze the following regimes 
\begin{itemize}
\item {\bf Slow Roll (SR)} \\
  This is the standard slow-roll  regime   with  $0<\epsilon \ll 1$,
  $\eta \ll 1$.
\item {\bf Super Slow Roll (SSR)}\\
In this this regime  $0<\epsilon \ll 1$, while $\eta$ is not
necessarily small and is taken   to be  constant.

\item {\bf $\pmb{w}$-Media (WM)}\\
  This is the  simplest inflationary regime and  corresponds to a
  constant $w$ taken in the range $-1 <w<-\frac{1}{3}$. Thus
  $\epsilon=3(w+1)/2$ is also constant and consequently $\eta=0$.
  From the background EMT conservation it follows that $c_s^2=w$,
  while the EFT gives in this case $c_b^2=w$.
\end{itemize}
For all regimes of interest, either the time variation of $\eta$ is
quadratic in the  small slow-roll parameters $\epsilon$, $\eta$ and
then negligible at the leading order or is strictly constant in  SSR.  Thus from
eq. (\ref{slow_param}), we get
\be
\epsilon=\epsilon_i \, \left(- H_{in} \, t\right)^{-\eta} \, , 
\ee 
which of course is valid also in the  WM  regime with $\eta=0$
and $\epsilon=$constant.
It is convenient to  parametrize  the scale
factor $a$ as follows
\be
a=\left(-H_{in} \, t\right)^\beta \, , \qquad {\cal H}=
\frac{\beta}{t} \, ,
\label{infpar}
\ee
where $\beta$ is directly related to the parameter $\epsilon$
\be
\beta=-\frac{1}{1-\epsilon_i}\, , 
\ee
and its time dependence can be  neglected in SR at the linear order,
while in the WM regime is strictly constant and in particular
$\beta= 2/\left(1+3w\right)$ .
In the SSR regime,  if $\eta <0$,  it is clear that $\epsilon$ quickly goes to zero and after the
inflation sets in, we can safely neglect it but not its time
derivative related to $\eta$.
\\
We point out that  with such a choice of the background behaviour $\epsilon$, $\eta$  and $c_L$  are 
the only independent parameters for  the  dynamics of an
isentropic medium    with  $\delta \sigma=0$.
In the more general case of an adiabatic medium,  also  $c_b^2$ comes into play; among other
things it  determines the dynamics of     the background
of  $\varphi^0$ according with 
\be
\bar\varphi''=(1-3\;c_b^2)\;{\cal H}\;\bar\varphi' \, ,
\ee
that, for   a constant  $c_b^2$ gives
\be
\bar\varphi'
=\bar\varphi_{in}'\;\left(\frac{a}{a_{in}}\right)^{(1-3\;c_b^2)} \, .
\ee

\subsection{Media  Inflation Models}

On a de Sitter (dS) background, where $\bar \rho+ \bar p\to 0$, $c_s$,
$c_L$ and $c_T$ become singular~\footnote{Note, however, that inflation takes place when $w<-\frac{1}{3}$ and strictly
speaking only slow-roll and the ultra slow-roll regimes need a
background close to dS where  $w\sim -1$.}, see
(\ref{cl}, \ref{cs}, \ref{cT}).
In order to prevent singularities that could
invalidate perturbation theory, a successful  inflationary phase
requires a careful choice of the Lagrangian $U$ describing the medium.
  In more detail, looking at the  expressions for the energy (\ref{ed})
density and pressure (\ref{pre}), for an adiabatic medium where $X$ and
the $\{y_n\}$ are absent, one has
\be
p+\rho=Y\, U_Y-b\,U_b-\frac{2}{3} \sum_{n=1}^3 n \,U_{\tau_n}  \tau_n\, .
\ee
A quasi-dS inflationary phase can be realised by taking a Lagrangian of the form
\footnote{The corresponding mass parameters $M_A$ generated by $U$ can be 
generically written as $M_A=M_{\Lambda_A}+(1+w) \;M_{V_A}$, where
$M_{\Lambda_A}$ is associated with $U^\Lambda$ while
$M_{V_A}$  with $V$.}
\be
U(\tau_i,\,Y)=\underbrace{U^\Lambda(\tau_i,\,Y)}_{ \propto
  M_{Pl}^2\;H^2} \;  \, +
\underbrace{V(\tau_i,\,Y)}_{ \propto (1+w) \, M_{Pl}^2\;H^2} \, ;
\ee
$U^\Lambda$ selects the dS background,
while the subleading part $V$ provides the small deviation from  it.
We have two  main options for the form of $U^\Lambda$:
\begin{itemize}
\item[({\bf a})]  $U^\Lambda=\Lambda$ is just a cosmological
  constant (CC).
\item[({\bf b})]  $U^\Lambda $ is a special case of the $\Lambda$-media~\cite{Celoria:2017idi} Lagrangians, 
 featuring  $w=-1$, with a non trivial dependence from the various operators.
\end{itemize}
The difference in the two options shows up in the values of the mass parameters $M_i$.
\\
$({\bf a})$ In the CC dominated case all masses are order $(1+w)$
\be
M_i\propto  (1+w) \;M_{V_i} \, 
\ee
and all various ``sound velocities'' $c_{s,\,L,\,T,\,b}^2$ are
parameters   given by $V$ and its derivatives. \\
For the $\L$-Media dominated case we can have $\L$-perfect fluids Lagrangian of the form $U^\L=U(b\;Y)$, 
isentropic $\L$-solids with $U^\L=U(Y\,\tau_1^{3/2},\,\frac{\tau_2}{\tau_1^2},\, \frac{\tau_3}{\tau_1^3})$ and special isentropic $\Lambda$-supersolids with a Lagrangian $U^\L=U( \frac{w_2}{w_1^2},\, \frac{w_3}{w_1^3})$.
For our purposes, the main feature of $\Lambda$-media is the
following set of relations among the mass parameters~\cite{Celoria:2017idi}
\be
M_{\Lambda_0}=M_{\Lambda_4},\qquad M_{\Lambda_2}=3\;(M_{\Lambda_3}-M_{\Lambda_4}) \, ;
\ee
for fluids $M_{\Lambda_2}=0$, thus in this case $M_{\Lambda_0}=M_{\Lambda_3}= M_{\Lambda_4}$.
Moreover, to avoid the above mentioned
singularities for the $c_i^2$, the mass parameters need to have the following structure
\be
M_2\propto (1+w)\;M_{V_2} \, \qquad \qquad 
M_{0,3,4}\propto M_\Lambda+ (1+w)\;M_{V_{0,3,4}} \,  ;
\label{Mstruct}
\ee
where $M_\Lambda$ is a unique common mass scale generated by
$U^\Lambda$. With the structure (\ref{Mstruct}), no singularity
related to $\rho+p \to 0$ will be present. This will be the case when
$U^\Lambda $ has the following features:
\begin{itemize}
\item The constraint on $M_2$ implies that the Lagrangian $U^\L$
  represents either a $\Lambda$-perfect fluid $U^\Lambda=
  U(b\;Y)$ or a specific massless $\Lambda$-solid for which
  $M_{\Lambda_2}=0$. One can take for instance $U(\frac{\tau_2}{\tau_1^2}-\frac{\tau_3}{\tau_1^3},Y)$.
\item The values of the parameter $c_b^2$ is driven by $U^\L$ and
  thus $c_b^2\sim -1 $, where the deviation from $-1$ is small and provided by
  $V$. The other parameters $c_{s,L,T}^2$ are determined by $V$.
\end{itemize}
Thus, at quadratic level, the main difference between the two
class of Lagrangians is the value of $c_b^2$, that, as we will see,
 is a key parameter for the entropic contributions to the power spectrum. For the
$\Lambda$-media dominated case $c_b^2$ is close $-1$, while it has generic value  in the CC dominated case.
In the WM regime, described in detail in section \ref{W-Media},
the background is not close to dS, $w$ is constant with $-1 < w=\text{const}
<-\frac{1}{3}$, realizing an accelerated expansion of the
Universe. The corresponding medium is a simple generalization of the
$\L$-medium (with $w=-1$) and the Lagrangian  is  given by eq. (\ref{W-M_lagrangian}).  

\subsection{The Comoving Curvature Perturbation \texorpdfstring{${\cal R}$}{Lg}}

Consider first the comoving curvature perturbation ${\cal R}$ whose
dynamics is described by (\ref{caneq}). For all the inflation regimes
under investigation, such equation assumes the form of a Bessel
equation plus an inhomogeneous term in the case of  
isentropic media proportional to entropy per particle perturbations; namely
\be
\label{caneq2}
{\cal R}_{c}{}''+ {\cal R}_{c}{} \left[ k^2 c_L^2+\frac{1}{t^2}\left(\frac{1}{4}-\nu^2 \right)\right]+ \Sigma_k (-H_{in} \, t)^\gamma=0 \,; 
\ee
where, by using the parametrization described in the previous section,
we have 
\be
\label{deltaSigma}
\begin{split}
& \nu^2=\frac{1}{4}\,\left(1-2\,\beta+\eta\right)^2-4\, c_T^2\, \beta^2 \, \epsilon_i\, ;\\[.2cm]
& \gamma=-3\, \beta \, c_b^2 +\frac{\eta}{2}\, ;\\[.2cm]
& \Sigma_k=\frac{\varphi'_{in}}{2\,\sqrt{6} \, c_L\, M_{pl}^2\,\sqrt{\epsilon_i}} \left[3\,c_b^2\, (c_b^2-c_s^2)+\epsilon_i \,(c_L^2-c_b^2)\right]\, \delta \sigma\, .
\end{split}
\ee
In order to compute the primordial power-spectrum, it is natural to take the   entropy per particle  perturbation $\delta \sigma$, as a classical external field.
\\
 Thus, no cross-correlation
terms $\left<\delta \sigma \,\,  {\cal R}^{(0)}\right>$ will be present. 
 The solution of eq. (\ref{caneq2}) can be written as 
 \be
 {\cal R}={\cal R}^{(0)}+{\cal R}^{(\delta \sigma)} \, ,
\ee
where ${\cal R}^{(0)}$ is the general solution of the
  homogeneous  part of
(\ref{caneq2}) obtained by setting  $\Sigma_k=0$, while ${\cal R}^{(\delta \sigma)}$
is a particular solution of (\ref{caneq2}).

\subsubsection{Isentropic solutions}

The solution of the homogeneous equation can be written as a linear combination of Hankel functions
\be
{\cal R}_c^{(0)}=\sqrt{-k\,t} \left[{\cal R}_1\; H^{(1)}_{\nu}(- c_L \, k\,t)+{\cal R}_2 \;H^{(2)}_{\nu}(- c_L \, k\,t)\right] \, .
\ee
Notice that our choice of the conformal time $t$ is such that $t \in
(-\infty, 0]$.
The scalar ${\cal R}$ is canonically quantized and the Bunch-Davies
is selected as the vacuum; this is equivalent to impose that for large $k$ (sub-horizon
mode) the mode is a suitable normalized  positive  frequency plane wave
$\exp(- i \,c_L\, t \, k)$, namely 
 \be
 {\cal R}_2=0 \, , \qquad |{\cal R}_1|^2 =  \frac{\pi}{4 \,M_{pl}^2\,k} \, .
 \ee
Thus, the mode assumes the form 
\be
\label{super_hor_R}
{\cal R}^{(0)}_k(t)=\frac{{\cal R}_c}{L_{\cal R}^{\frac{1}{2}}}=  \left(\frac{-c_L^2\, \pi}{8 \, M_{pl}^2 \, \epsilon_i \, H_{in}^{2\beta-\eta}}\right)^{\frac{1}{2}}(-t)^{\frac{1}{2}-\beta+\frac{\eta}{2}} \,\,H_{\nu}^{(1)} (- c_L \, k\,t) \,  ,
\ee
and the power spectrum of ${\cal R}$, defined in terms of the two-point
function of the associated free field, 
will be given by
\be
{\cal P}_{{\cal R}^{(0)}}= \frac{k^3}{2 \pi^2}\,\left|{\cal R}^{(0)}_k(t)\right|^2 \, .
\ee
In practice, the power spectrum is important on superhorizon scales, namely
in the limit $|k\, t| \ll 1 $. Depending on chosen sign for $\nu$ in the
Hankel function, we have that
\be
\left.H_{\nu}^{(1)}(- c_L \, k\,t)\right|_{\nu >0}\sim (k\,t)^{-\nu}\, , \qquad 
\left.H_{\nu}^{(1)}(- c_L \, k\,t)\right|_{\nu <0}\sim (k\,t)^{\nu} \, .
\ee  
Of course, the final result will be independent from the choice of sign
for $\nu$  being relation (\ref{caneq2}) symmetric. Hereafter we always choose $\nu$ to be positive. 
Thus, in the superhorizon limit we get for the power spectrum
\be
\label{power_spectrum}
{\cal P}_{{\cal R}^{(0)}}= \frac{2^{2 \,\nu-4}\, c_L^{2-2\,\nu}\, \Gamma(\nu)^2\,(H_{in})^{-2\,\beta+\eta} }{\epsilon _i\, \pi^3 \, M_{pl}^2}\,(-t)^{1-2\,\nu-2\,\beta+\eta}\, k^{n_s-1} \,.
\ee
Traditionally the power spectrum ${\cal P}$ for a generic
scalar  quantity is split in an amplitude $A$ and a tilt parameter $n_s$,
according with
\be
{\cal P} = A \, k^{n_s-1} \, .
\label{split}
\ee
From (\ref{power_spectrum}), one can read off the tilt parameter
for ${\cal R}^{(0)}$ which is given by $n_s=4-2\,\nu$. Regardless of
the inflationary regime choice, 
$n_s$ close to one, requires $\nu \approx \frac{3}{2}$.
As we will see in section \ref{viol}, a generic important feature  is
that, on superhorizon scales, the power spectrum will still have a
residual time dependence.

\subsubsection{Entropic modes}

Let us study the effect of the entropic perturbations in the
power spectrum. The particular solution of eq. (\ref{caneq2}) can be written as
\be
 {\cal R}_c^{(\delta \sigma)}= \frac{1}{\nu}\, H_{in}^\gamma\, \Sigma_k \, (-t)^{2+\gamma} \left[{\cal F}\left(\alpha,\, \nu,\, c_L^2,\, k\,t\right)-{\cal F}\left(\alpha+\nu,\, -\nu,\, c_L^2,\, k\,t\right)\right] \, ,
\ee   
where $\alpha=\frac{1}{4}\left(3+2\,\gamma-2\,\nu\right)$, and the
function ${\cal F}$ is a combination of regularized confluent
hypergeometric and regularized generalized hypergeometric functions
\be
{\cal F}\left(\alpha,\, \nu,\, c_L^2,\, k\,t\right) = -\frac{1}{4\,\alpha} {\cal F}_0 \left(1+\nu,\,-\frac{1}{4}\,c_L^2\, (k\,t)^2\right){\cal F}_1 \left(\alpha, \,  \{1-\nu,\, \alpha+1\},-\frac{1}{4}\, c_L^2\, (k\,t)^2\right)\, .
\ee
 At small scales (sub horizon) the entropic contribution ${\cal
   R}^{(\delta \sigma)}$ to ${\cal R}$ has the form
\be
{\cal R}^{(\delta \sigma)} \sim \frac{\Sigma_k}{\sqrt{\epsilon_i}} \, (-k\,t)^{-(1+3\, c_b^2)\,\beta+\eta} \qquad \qquad |k\,t| \gg 1 \, .
\ee
In order to not alter the Bunch-Davies vacuum, such contribution must
be subdominant at small scales and early time; this will be the case when
\be
\eta  -(1+3\,c_b^2)\,\beta<0 \;\;\Rightarrow  \;\;c_b^2<\frac{\eta-\beta}{3\,\beta}\, . 
  \label{BDC}
\ee
With no particular assumption 
on the dependence of $\delta \sigma$ on $k$, the above condition is sufficient. However,
(\ref{BDC}) could be too stringent   if $\delta
\sigma$ is a function of $k$ with a particular form. This might be the case when $\delta
\sigma$   produces the seed for primordial black holes and then it is
 peeked on a momentum $k_0$ so large that  cannot be
directly tested and condition (\ref{BDC}) does not necessarily apply.

Provided that the vacuum state is not altered, at large scales we get the following expression for the entropic contribution to the
curvature perturbation
\be
{\cal R}^{(\delta \sigma)}=-\frac{4\, \sqrt{3}\,c_L\,
  H_{in}^{-\beta+\gamma+\frac{\eta}{2}} \Sigma_k}{\left[(2\, \gamma
    +3)^2-4\, \nu ^2\right]\sqrt{2\, \epsilon_i}} (-t)^{2+\gamma-\beta+\frac{\eta}{2}} \qquad \qquad  |k\,t| \ll 1\, . 
\ee
The entropic mode $\delta \sigma$ is a classical source, 
and the total power spectrum ${\cal P}_{\cal R}={\cal P}_{{\cal R}^{(0)}}+{\cal P}_{{\cal
    R}^{(\delta \sigma)}}\ $ is simply given by the sum of
(\ref{power_spectrum}) and
\be
\label{spectrum-deltasigma}
{\cal P}_{{\cal R}^{(\delta \sigma)}}= \frac{12\, c_L^2 \,H_{in}^{-2\,\beta+2\,\gamma+\eta} }{\pi^2 \left[(2\,\gamma+3)^2-4\,\nu^2\right]^2\, \epsilon_i} \,|\Sigma_k|^2\;k^3\;(-t)^{4+2\,\gamma-2\,\beta+\eta} \, .
\ee
Note that the time dependence of ${\cal P}_{{\cal R}^{(\delta \sigma)}}$
can be neglected when $c_b^2=c_s^2$ i.e. for different scenario it happens when
   \begin{itemize}
   \item SR: $c_b^2\approx-1$.
   \item SSR: $c_b^2=-1-\frac{\eta}{3}$.
   \item In WM it is always exactly constant being $c_b^2=w=c_s^2$.
   \end{itemize}
From the relation (\ref{deltaSigma})~\footnote{When $c_b^2=c_s^2$, not
  only the entropic part of the power spectrum is time independent,
  but also $\Sigma_k$ is of order $\sqrt{\epsilon_i}\,\delta\sigma$,
  instead of $\frac{\delta\sigma}{\sqrt{\epsilon_i}}$.}, one can check
that the entropic
contribution to the total power spectrum is always of order
$\frac{1}{{\epsilon_i}^2}|\delta \sigma|^2$. The contribution of
$\delta \sigma$ to the total power spectrum can be used to get 
the enhancement required to form primordial black holes that could
serve as a MACHO-like dark matter
candidates~\cite{Bird:2016dcv,Sasaki:2016jop,Clesse:2016ajp,Espinosa:2017sgp},
see for instance~\cite{Ballesteros:2018wlw} for a recent discussion.\\
The various inflationary regimes are summarised in table~\ref{VV}.
\begin{table}[htp]
\begin{center}
\begin{tabular}{|c|c|c|}
\hline
SR& SSR& WM\\\hline
$\epsilon_i,\,\eta\ll1$& $\epsilon_i=0,\,\eta\sim{\cal O}(1)$ &$\epsilon_i\sim{\cal O}(1),\,\eta=0$\\
\hline
$\beta=-1-\epsilon_i$ & $\beta=-1$ & $ \beta=-\frac{1}{1-\epsilon_i}$\\
\hline
$c_s^2=-1+\frac{2}{3}\epsilon_i-\frac{\eta}{3}$ & $c_s^2=-1-\frac{\eta}{3}$ & $c_s^2=-1+\frac{2\;\epsilon_i}{3}$
\\
\hline
$\nu=\frac{3}{2}+\frac{\eta}{2}- c_L^2\;\epsilon_i$
&
$\nu=\frac{3+\eta}{2}$
&
$\nu=\frac{3}{2} \sqrt{1-\frac{8\; c_L^2\;
   (1+w)}{\left(1+3\,w\right){}^2}}$\\
\hline
$\gamma=3(1+\epsilon_i)\;c_b^2+\frac{\eta}{2}$ &$\gamma=3\;c_b^2+\frac{\eta}{2}$ &$\gamma=\frac{3\;c_b^2}{1-\epsilon_i}$ \\
\hline
 \end{tabular}
\end{center}
\caption{Overview of the considered inflationary
  regimes.}
\label{VV}
\end{table}%

\subsection{The Uniform Curvature Perturbation \texorpdfstring{$\zeta$}{Lg}}

\label{entro}

At superhorizon scales, in the standard single field slow-roll
inflation,  comoving and uniform curvature perturbations
  are conserved and equivalent ${\cal R}= \zeta$  at
  superhorizon scales.
This is the case because the Weinberg
theorem holds. However, in general this will be not the case and  an independent study of the dynamics of $\zeta$ is required.
 Unfortunately, though the
dynamical equation for $\zeta$ is similar in form to (\ref{caneq}),
see (\ref{zetaeq}), even in the single field inflation is much harder to solve.
However, once ${\cal R}$ is known, one can find $\zeta$ by
using the general relation  (\ref{Rzetarel}). As for ${\cal R}$ we can
split $\zeta$ into an adiabatic part $\zeta^{(0)}$ and an entropic
contribution $\zeta^{(\delta \sigma )}$ originating from the presence
of $\delta \sigma$; thus   ($\zeta=\zeta^{(0)}+\zeta^{(\delta \sigma )}$) 
\be
\begin{split}
& \zeta^{(0)}={\cal R}^{(0)}\, \frac{c_s^2}{c_L^2}-\frac{{\cal R}^{(0)}{}'}{3 \,{\cal H} \,c_L^2}\, ,\\ 
& \zeta^{(\delta \sigma)}={\cal R}^{(\delta \sigma)} \,\frac{c_s^2}{c_L^2}-\frac{{\cal R}^{(\delta \sigma)}{}'}{3\, {\cal H} \,c_L^2}+\frac{(c_b^2-c_L^2)\, H_{in}^{-\beta+\gamma+\frac{\eta}{2}}\, \Sigma_k}{\sqrt{6} \,\sqrt{\epsilon_i}\, c_L \, \beta^2 \, \left[3\, c_b^2 \,(c_b^2-c_s^2)+\epsilon_i \,(c_L^2-c_b^2)\right]} (-t)^{2+\gamma-\beta+\frac{\eta}{2}}\, .
\end{split} 
\ee
From eq. (\ref{super_hor_R}), the adiabatic part of $\zeta$ can be written as
\be
\zeta^{(0)}=\sqrt{\frac{\pi \, t}{2\, \epsilon_i}}\left(-t H_{\text{in}}\right){}^{\frac{\eta }{2}-\beta } \left[\mathcal{A}\; H_{\nu }^{(1)}(-c_L \,k\, t)+
   \mathcal{B} \; k \,t\, H_{\nu -1}^{(1)}(-c_L\, k \,t)\right]\, ,
\ee
where 
\be 
{\cal A}=\frac{2\,\nu-1-\eta+2\,\beta\, (1+3\, c_s^2)}{12 \, c_L \,
  \beta \, M_{pl}} \,, \qquad {\cal B}=\frac{1}{6\, \beta\,
  M_{\text{pl}}} \, .
\label{constdef}
\ee
While ${\cal B}$ will be always different from zero, this is not the
case for  ${\cal A}$.   We have that 
\begin{enumerate}
\item SR: ${\cal A}=-\frac{1}{2\,c_L\;M_{pl}}$; 
\item SSR: ${\cal A}=0$ \footnote{   The fact that $\zeta$ is a single
    Hankel function suggests that    the dynamic of
    $\zeta_c$  can be described by  an equation of the form  (\ref{caneq2}).}; 
\item WM: ${\cal A}= \frac{1+3\,w}{8\,c_L^2\,M_{pl}}\,\left(1+\,\sqrt{1-\frac{8\,c_L^2\;( 1+w)}{(1+3\,w)^2}}\right) $.
\end{enumerate}
The knowledge of  ${\cal A}$  is crucial to get the correct
superhorizon limit;   indeed  for  $-k\, t \ll 1 $, we have 
\be
\label{Zeta-exp}
\zeta \sim {\cal A} \, (-k\,t)^{-\nu} \left[...\right]+\left({\cal A}\, c_L
  -2\, {\cal B}\right) (-k\,t)^{-\nu+2} \left[...\right] \, .
\ee  
Thus, in the SSR regime, being ${\cal A}=0$, only
the Hankel function $H_{\nu -1}^{(1)}$  contributes to the power
spectrum and one needs to expand such function to the next to leading order. 
By using the previous results, one can determine the general form of
the $\zeta$ power spectrum 
\be
\begin{split}
\label{zeta-power-spectrum}
& {\cal A}=0\,:\; {\cal P}_{\zeta^{(0)}}=\frac{2^{2(\nu-2)}\,{\cal B}^2\, c_L^{2(1-\nu)}\, H_{in}^{\eta-2 \beta} \Gamma \left(\nu-1\right)^2}{\pi^2 \epsilon_i}\,(-t)^{5-2\,\beta+\eta-2\,\nu} \, k^{7-2\,\nu}  \, , \\
&{\cal A}\neq 0\,: \; {\cal P}_{\zeta^{(0)}}= \frac{2^{2 (\nu -1)} \,\Gamma (\nu )^2 \,c_L{}^{-2 \nu} \,H_{\text{in}}^{\eta -2 \beta } 
   \mathcal{A}^2}{\pi ^3 \,\epsilon_i} \, (-t)^{1-2 \,\beta +\eta -2\, \nu }\, k^{3-2\, \nu }\,.
\end{split} 
\ee
The origin of ${\cal A}=0$ in the SSR  is  a peculiar  background
dynamics which leads to a dramatic violation of the Weinberg theorem;
besides the residual time evolution of superhorizon modes, the
tilt parameters of  ${\cal P}_{\zeta}$ and ${\cal P}_{{\cal R}}$ are very
different; namely 
\be
n_{s\,{\cal R}}=4- 2 \,\nu \, , \qquad n_{s\,\zeta}=4 +n_{s \, {\cal R}} =8- 2 \,\nu\, .
\ee
In the case of SR and WM inflation 
 $n_s=n_{s\,{ \scriptscriptstyle \cal R}}=n_{s\, \scriptscriptstyle\zeta}$. 
 \\
 Thus, in the SSR
 limit, the post inflationary evolution needs a close scrutiny to
 decide how the universe, after the reheating, settles into the
 ``standard'' hot phase. This will be analyzed in detail in section \ref{Post-Inf}.\\
 For what concerns the entropic contribution to the $\zeta$ curvature perturbation, at superhorizon scales $\zeta^{(\delta \sigma)}$  is
proportional to  ${\cal R}^{(\delta\sigma)}$
\be
\label{prop}
\zeta^{(\delta \sigma)} \sim \frac{\Sigma_k\,
  H_{in}^{\gamma-\beta+\frac{\eta}{2}}}{\sqrt \epsilon_i }(-t)^{\gamma+2-\beta+\frac{\eta}{2}}
\, , \qquad \zeta^{(\delta \sigma)} = {\cal C} \; {\cal
  R}^{(\delta\sigma)} \, ;
\ee
the parameter ${\cal C}$ is given by
\be
{\cal C}=\frac{(1+3 \, c_s^2)}{3 \,c_L^2}- \frac{1}{\beta^2}
\frac{(c_b^2-c_L^2)}{12 \, c_L^2} \frac{\left[(3+2 \, \gamma)^2-4
    \nu^2\right]}{\left[3 \, c_b^2\, (c_b^2-c_s^2)+\epsilon_i \,(c_L^2
    -c_b^2)\right]}-\frac{(4+2 \, \gamma+\eta)}{6 \, c_L^2 \, \beta} \, .
\ee
Thus
\be
\label{zeta_delta_sigma}
{\cal P}_{\zeta^{(\delta \sigma)}}= |{\cal C}|^2 \,{\cal P}_{{\cal R}^{(\delta \sigma)}} \,.
\ee
In particular, in the
three regimes of interest, the constant ${\cal C}$ is close or equal
to one; namely
\be
\label{C_one}
{\cal C}_{SSR}={\cal C}_{WM}=1\, , \qquad {\cal C}_{SR}=1+O(\epsilon_i^2, \, \epsilon_i \, \eta)\, .
\ee
The bottom line is that  ${\cal P}_{\zeta^{(\delta\sigma)}}$ and
${\cal P}_{{\cal R}^{(\delta\sigma)}}$  are the same.
Hereafter we will use the symbol ${\cal P}^{(\delta \sigma)}$ for both ${\cal P}_{{\cal R}^{(\delta \sigma)}}$ and ${\cal P}_{\zeta^{(\delta \sigma)}}$.

\subsection{Violation of the Weinberg Theorem}
\label{viol}
From the above discussion it is clear that the Weinberg
theorem is violated for a generic adiabatic media.
For solids,  the anisotropic
stress does not vanish for small $k$ and in general, for
non-barotropic media, perturbations are not purely adiabatic. In 
addition, in the SSR, the would-be decreasing mode becomes the
dominant one. The violation manifests itself as an ``anomalous''
behaviour of both $\zeta$ and ${\cal R}$ that will be different and
not conserved at superhorizon scales, affecting the scalar power spectrum.
In the isentropic case ($\delta \sigma=0$), from (\ref{power_spectrum}) and (\ref{zeta-power-spectrum}), it
follows that the time dependence of the amplitude   in the ${\cal R}$ power 
spectrum has the following form 
\be
A_{\cal R} \sim (-t)^{1-2\,\beta+\eta-2\, \nu}\, ;
\ee
  while  for $\zeta$ 
\be
 A_{\zeta} \sim \begin{cases} 
(-t)^{1-2\,\beta+\eta-2 \,\nu}   & {\cal A} \neq 0 \, ; \\
 (-t)^{5-2\,\beta+\eta-2\, \nu} & {\cal A} = 0 \, .
\end{cases}
\ee
When ${\cal A} \neq 0$ (see definition (\ref{constdef})), the power
spectra of $\zeta$ and ${\cal R}$ have the same time dependence though they are not
identical; indeed
\be
\frac{{\cal P}_{\zeta^{(0)}}}{{\cal P}_{{\cal R}^{(0)}}} =\begin{cases}  
 4\, {\cal A}^2 \frac{M_{pl}^2}{c_L^2}  \\
 M_{pl}^2 \, {\cal B}^2
 \,\frac{\Gamma(\nu-1)^2}{\Gamma(\nu)^2}(-k\,t)^4 
\end{cases}
=
\begin{cases}
 \frac{\left[2\, \nu-1-\eta+2 \,
     \beta\,(1+ 3 \, c_s^2)\right]^2}{36 \, c_L^4\,
\beta^2} &  {\cal A} \neq 0 \,; \\
\frac{1}{36\,
  \beta^2}\frac{(-k\,t)^4}{(\nu-1)^2} & {\cal A} = 0
\end{cases} \, .
\label{specrel}
\ee 
Taking into account the explicit values of the parameters in the three
inflationary regimes, we can summarize the effects of the violation of
Weinberg theorem as follows: 
\begin{itemize}
\item SSR: ${\cal A}= 0$, both time dependence and amplitudes are
  different. At superhorizon scales, ${\cal P}_{\zeta^{(0)}}$ is much
  smaller than ${\cal P}_{{\cal R}^{(0)}}$ and suppressed by a factor $(k\,t)^4$; 
\item SR: ${\cal A}\neq 0$; both $\zeta$ and ${\cal R}$  grow in time
  in the same way at superhorizon scales however the
  amplitudes are different. When $n_s-1 \approx 0$, both spectra are
  almost flat and the time dependence is negligible as in single field inflation.
\item WM: ${\cal A}\neq 0$; both $\zeta$ and ${\cal R}$ grow in time
  in the same way at superhorizon scales however the amplitudes are different. 
 The time dependence can be very different from the one in the SR.
\end{itemize}
The detailed analysis for the different media is given in the following section.

\section{Media Inflation}
\label{Media-inflation}
In this section we  will study the
relevant inflationary regimes for each class of media
  as special cases  of  the general analysis of the dynamics of
${\cal R}$ and $\zeta$.

\subsection{Extended Fluid Inflation }

\label{Fluid_inf}
 We  will extend the analysis of~\cite{Chen:2013kta}
by considering a general fluid with a non-barotropic equation of state  which supports
entropy perturbations; in addition we will study in detail the differences in the  dynamics of
${\cal R}$ and $\zeta$ .\\
The main feature of a fluid is the absence of anisotropic stress; in
our language $M_2=0$ and thus we have that $c_L^2=c_{s }^2$.
From (\ref{speed}), it is clear that the condition (\ref{pos}) is
incompatible with an inflationary SR or WM regimes. Nevertheless, the
super slow-roll regime is possible, taking $\eta <-3$\footnote{Only if $\eta <-3$ the sound speed is positive}. As discussed
before, in the SSR regime we can set $\epsilon=0$ and we have that
\be
\nu=-\frac{(\eta+3)}{2}  \, , \qquad \beta=-1 \, , \qquad \gamma= 3 \,
c_b^2+ \frac{\eta}{2} \, .
\ee
 On superhorizon scales, our general analysis gives in the  isentropic case
 \be
{\cal R}^{(0)} = C_1 + C_2 (-k \,t)^{\eta+3}  \, , \qquad \qquad \zeta^{(0)} =
C_1 + C_2 \, (-k \,t)^{\eta+5} \qquad  |k \, t| \ll 1 \, .
\ee
As explained in section \ref{Toy_model}, though all perturbations are adiabatic, the very fast
decrease of $\epsilon$ in the super slow-roll regime turns the
decreasing mode in ${\cal R}$ and $\zeta$ in a growing one when $\eta
< -3$. The constant mode is still present but sub-leading, in this
sense the Weinberg theorem is violated. 
In this case ${\cal A}$ is zero and by using  (\ref{power_spectrum})
and (\ref{specrel}), we get the power spectra
\be
\begin{split}
& {\cal P}_{{\cal R}^{(0)}}=\frac{2^{-(7+\eta)}\,\Gamma\left(-\frac{(3+\eta)}{2}\right)^2\, c_s^{5+\eta}\, H_{in}^{2+\eta}}{\pi^3\, M_{pl}^2\, \epsilon_i}\; t^{6+2\eta} \; k^{6+\eta} \, , \\
& {\cal P}_{{\zeta}^{(0)}}=\frac{1}{9}\frac{(k\, t)^4}{(\eta+5)^2}\; {\cal P}_{{\cal R}^{(0)}}\,.
\end{split}
\ee
One of the peculiar feature of fluid inflation is that the requirement
of a scale-free spectrum for $\zeta$ and ${\cal R}$, e.g. $n_s=1$,
gives very different values of $\eta$. Indeed,
\begin{itemize}
\item $n_s=1$ for ${\cal R}$
gives $\eta=-6$ with $c_s^2=1$, 
\item   $n_s=1$ for $\zeta$
implies $\eta=-10$ and $c_s^2=7/3>1$.
\end{itemize}
Leaving aside the possible
Cherenkov decay due to $c_s^2 >1$, whatever choice is made $A_{\zeta}
\neq A_{\cal R}$ and the spectrum cannot be simultaneously scale-free.
As it will discussed in section \ref{Post-Inf}, $\zeta$ is
continuous in the approximation of an instantaneous reheating taking
place on the $\rho=$ constant hypersurface, while ${\cal R}$ 
jumps to align with the value of $\zeta$. Thus, the spectrum which
needs to be scale free is the one of $\zeta$ which selects  $\eta=-10$
and the sound speed will be quite far from being sub-luminal,
in contrast to the result in~\cite{Chen:2013kta}.  

The case of a non-barotropic fluid is described by the Lagrangian
$U(b,Y)$; thus the power spectrum $ {\cal P}_{{\cal R}^{(0)}}$ is
supplemented by an entropic contribution.
Specializing  (\ref{deltaSigma}) and  (\ref{spectrum-deltasigma}) to the super slow-roll regime, we get 
\be 
{\cal P}^{(\delta \sigma)} \propto \frac{k^3 \,|\Sigma_k|^2}{\epsilon_i} (-t)^{6\,(1+c_b^2)+2\,\eta} \,. 
\ee
The new key parameter is $c_b^2$ that controls the amount of
non-adiabatic perturbation present and also, together with $\eta$, the
superhorizon behaviour; namely
\be
\frac{{\cal P}^{(\delta\sigma)}}{{\cal P}_{\zeta^{(0)}}} \sim
k^{-7-\eta}\, |\Sigma_{k}|^2 \,t^{6(1+c_b^2)-10}  \qquad \qquad | k \, t |
\ll 1 \, . 
\ee
For super slow-roll, condition (\ref{BDC}) becomes 
\be 
c_b^2 < -\frac{(1+\eta)}{3}\,;
\ee
in particular, if a flat spectrum requires $\eta=-10$, thus the condition reduces to $c_b^2 < 3$. 

Concluding, if $2/3 < c_b^2 < 3$ then ${\cal P}^{(\delta \sigma)} \ll {\cal
  P}_{\zeta}^{(0)}$ and the entropic contribution is irrelevant; while
if $c_b^2 < 2/3$, then ${\cal P}^{(\delta \sigma)} \gg {\cal
  P}_{\zeta}^{(0)}$ and the entropic contribution dominates at
superhorizon scales. Particularly interesting will be the case $c_b^2 = 2/3$, 
where the two contributions are comparable.

\subsection{Extended Solid Inflation}

As a second large class of media we consider finite temperature
solids~\cite{Celoria:2017bbh} described by the Lagrangian
$U(Y,\tau_1,\tau_2,\tau_3)$, a generalization of zero temperature
solids~\cite{Endlich:2012pz} where the operator $Y$ and
thus the scalar field $\varphi^0$ were absent. Contrary to fluids,
solids have anisotropic stress (proportional to $M_2$) which changes the propagation velocity of the
scalar modes from $c_s^2$ to $c_L^2$, see (\ref{quadact}), and
the speed of propagation of vector modes is the transverse speed $c_T$.

\subsection{Slow-Roll Inflation for Solids}
The presence of an anisotropic stress,
 opens up for solids the possibility of slow-roll inflation with
$\epsilon$ and $\eta$ small and $c_L^2>0$. 
As already discussed in section \ref{Toy_model},  the presence of an
anisotropic scalar $\Xi=2 M_{pl}^2 (\phi-\psi)$ turns on the mass term
$M_2$ in the dynamical eq. (\ref{caneq}) for ${\cal R}$ and we expect a 
violation of the Weinberg theorem.\\
Making no assumption on $c_T^2$,  at the leading order in
the slow-roll parameters $\eta$ and $\epsilon$, one gets
\be
\beta=-1-\epsilon_i \, , \qquad \gamma=3\, c_b^2
\,(1+\epsilon_i)+\frac{\eta}{2}\, , \qquad \nu=
\frac{3}{2}+\epsilon_i+\frac{\eta}{2}-\frac{4}{3} \,c_T^2\, \epsilon_i
\,=\frac{3}{2} +\frac{\eta}{2}-\,c_L^2\, \epsilon_i  \,.
\ee
The violation of the Weinberg theorem can be directly inferred by the
asymptotic behaviour of ${\cal R}$ given by
\be
{\cal R} \propto (-t)^{\frac{1}{2}-\beta+\frac{\eta}{2}-\nu}=(-k\,t)^{\frac{4}{3} \,c_T^2\,
  \epsilon_i}\approx 1-\frac{4}{3}\,\epsilon_i\,c_T^2\, \log(- k\,t)  \qquad
\qquad | k \, t| \ll 1\, ;
\label{logrow}
\ee
where $\frac{1}{2}-\beta+\frac{\eta}{2} \neq \nu$ and the difference can be linked to
the presence of a non-negligible transverse speed $c_T$, or
equivalently to a mass parameter $M_2$. Modulo a logarithmic growing
correction of order $\epsilon$, ${\cal R}$ is constant in the superhorizon limit. Using eq. (\ref{power_spectrum}), the amplitude at the leading order in slow-roll parameters is given by
\be
{\cal P}_{{\cal R}^{(0)}}= A_{\cal R}  \, k^{3 -2 \, \nu }\, , \qquad  A_{\cal R} = \frac{H_{in}^{-2\beta+\eta}}{8 \pi ^2 \,c_L\,  \epsilon_i \, M_{pl}^2} (-t)^{\frac{8}{3} \, c_T^2 \, \epsilon_i} \, .
\label{PR1}
\ee
In the power spectrum, the violation of the Weinberg theorem manifests
 in log-like time dependence  due to the presence of the factor $
a^{-\frac{8}{3} c_T^2\,\epsilon_i }$.
Let us now come to $\zeta$.
Using  (\ref{specrel}) and (\ref{zeta-power-spectrum}) with ${\cal A }\neq 0$ and $\nu > 0$, on superhorizon scales we get
\be
\zeta^{(0)}
 = {\cal R}^{(0)} \, \left[\frac{c_s^2}{c_L^2}-\frac{4}{9} \frac{c_T^2}{c_L^2}
   \, \epsilon\right] +O( kt)  \, , \qquad {\cal P}_{\zeta^{(0)}}= A_{\zeta} k^{3 -2 \, \nu } \,, \qquad \frac{A_{\zeta}}{A_{\cal R}}=\frac{1}{c_L^4} \, .
 \label{zetaise}
\ee
Therefore, the spectral index $n_s$ is the same of ${\cal R}$ but the
amplitude is different. Such a feature, which has its origin in the
violation of the Weinberg theorem, was present also in the case of
fluid inflation, however here is mitigated by the slow-roll dynamics
in which $\eta$ and $\epsilon$ are both small.\\
Now let us extent our analysis to the more general case of an adiabatic media for
which $\delta p \neq  c_s^2 \, \delta \rho$ extending the result in~\cite{Endlich:2012pz}. 
The additional contribution to the scalar power spectrum comes from the
correlation of $\left<\delta \sigma \, \delta\sigma\right>$, and on
superhorizon scales we have
\be
\label{Solid_dSigma_spectrum}
 {\cal P}^{(\delta \sigma)}=\frac{12\, c_L^2 \, H_{in}^{-2\, \beta+\eta}\, |\Sigma_k|^2 \, k^3}{ \pi^2\left \{ \left[6\,c_b^2 (\epsilon_i +1)+\eta
       +3\right]^2-4 \, \nu^2\right \}^2 \epsilon_i}\,(-t)^{6\,( 1+c_b^2 )+2\,\epsilon_i (1+3\,
     c_b^2)+2\,\eta} \, .
\ee
At the leading order in the amplitude
\be
\frac{{\cal P}_{\zeta^{(0)}}}{{\cal P}_{{\cal R}^{(0)}}}=
\left(\frac{c_s^2}{c_L^2}\right)^2 \, , \qquad \frac{{\cal
    P}_{\zeta^{(\delta \sigma)}}}{{\cal P}_{{\cal R}^{(\delta \sigma)}}}=1 \, .
\label{enratio}
\ee
As done for fluids, in presence of entropic modes,  it is important
to determine the relative size of the two contributions in the power
spectrum. Let us consider for instance, the $\zeta$ power spectrum; at
the leading order in the slow-roll parameters we have
\be
\frac{{\cal P}^{(\delta\sigma)}}{{\cal P}_{\zeta^{(0)}}} 
=  \frac{96  \, c_L^7 \, M_{pl}^2\,|\Sigma_k|^2 \, k^{2\,\nu}}{ \left \{ \left[6\,c_b^2 \,(\epsilon_i +1)+\eta
       +3\,\right]^2-4 \, \nu^2\right \}^2 }\,(-t)^{6\,( 1+c_b^2 )+2 \,\epsilon_i(3\,c_b^2-c_L^2)+2\,\eta} \, .
\ee   
Condition ($\ref{BDC}$), neglecting linear corrections in slow-roll
 parameters, requires that $c_b^2<-\frac{1}{3}$, thus as in the SSR regime
  we can distinguish three different regimes.
\begin{enumerate}
\item $- 1 < c_b^2 <-\frac{1}{3}$\\
  In this case the denominator of
  the previous relation is of order one and $|\Sigma_k|^2 \sim |\delta
  \sigma|^2/ \epsilon_i $, while the time dependence factor kills the
  ratio at late times. The entropic contribution is irrelevant.
\item $c_b^2=-1$\\
  The time dependence of the two
  contributions drops off in the ratio. From eq. (\ref{deltaSigma}), it follows
  that $|\Sigma_k|^2 \sim \epsilon_i |\delta \sigma|^2$ and then 
 the ratio is of order $|\delta\sigma^2|/\epsilon_i$. Thus, only if
$\delta\sigma$ is parametrically much smaller then
$\sqrt{\epsilon_i}$, the entropic contribution will be sub-dominant.
\item $c_b^2<-1$\\
  The ratio is again of order $|\delta \sigma|^2 /\epsilon_i$; however
  the time dependent factor grows at late time and, unless
  $\delta\sigma$ is tiny, the entropic contribution will be the
  dominant one.
\end{enumerate}
Of course also for solids one can consider a super slow-roll
regime; results of the fluid case can be extended by replacing $c_s^2$
with $c_L^2$ emphasizing the super-luminal problem 
in the case of a scale free uniform power spectrum.

\subsection{$w$-Media Inflation}
\label{W-Media}
\begin{figure}[ht!]
\begin{center}
\includegraphics[width=8cm]{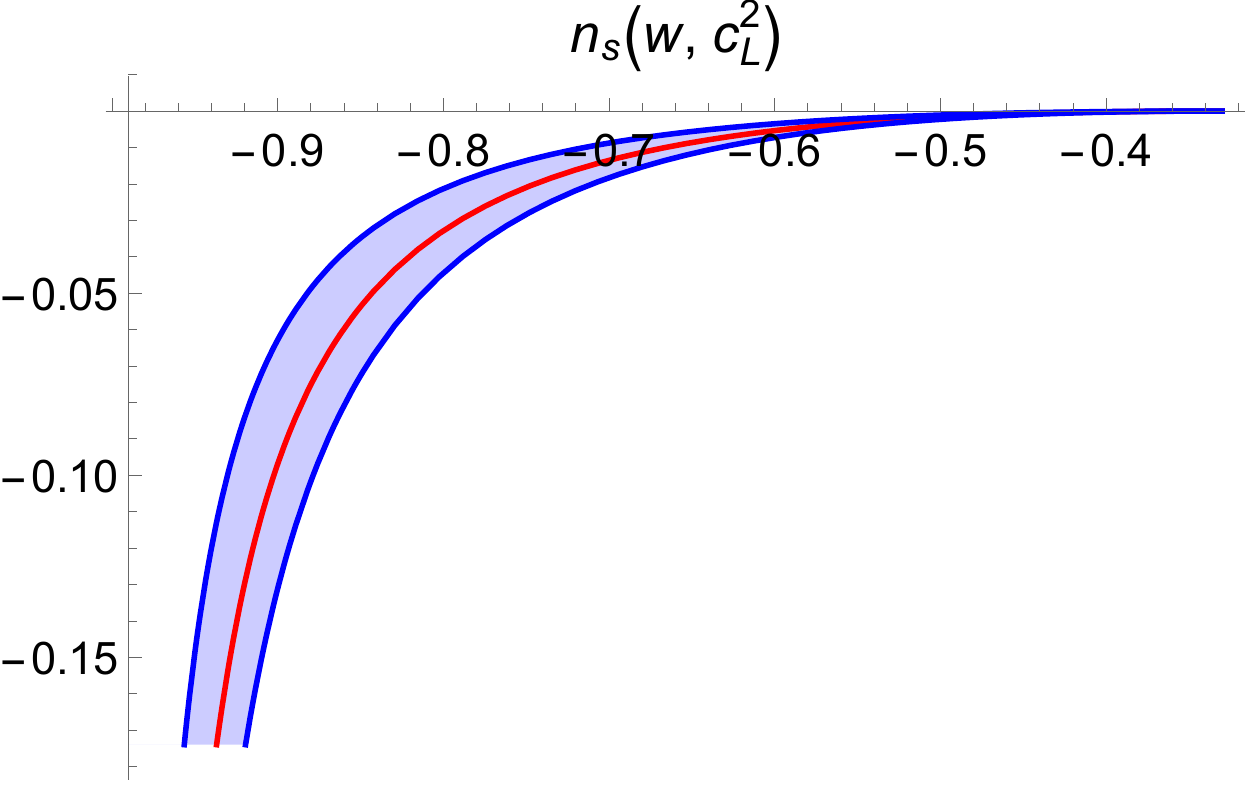}
\caption{Tilt parameter $n_s$ contour lines as a function of $w$ and $c_L^2$ parameters, using the experimental constraint $n_s=0.96 \pm 0.014$ with CL $95\%$ in the case of no tensor perturbations and no momentum-dependence of the tilt.}
\label{ns}
\end{center}
\end{figure}

The $w$-Media regime is characterized by $c_b^2=c_s^2=w$, with 
\be
-1< w<-\frac{1}{3} .
\ee 
Being the velocity of scalar modes for a solid $c_L$ (positive in general), for such a media an inflationary phase is possible.
The most general adiabatic $w$-Media can be described by the following Lagrangian
\be
\label{W-M_lagrangian}
U=\tau_1^{\frac{3(1+w)}{2}}V\, \left(\frac{Y}{\tau_1^{\frac{3\, w}{2}}},\,\frac{\tau_2}{\tau_1^2}, \, \frac{\tau_3}{\tau_1^3}\right) \, . 
\ee
and for them
\be
\varphi'=\varphi'_{in} \,(-H_{in}\, t)^{\frac{2(1-3\,w)}{1+3\,w}} \, ,\qquad M_i= m_i \left(-H_{in}\,t\right)^{\frac{2(1-3\,w)}{1+3\,w}} , \qquad m_4= -w\, m_0 \, , \qquad m_1=0 \, ,
\ee
where the time independent parameters $m_i$ are functions of the   1st and 2nd Lagrangian derivatives with respect to the $Y$ and $\tau_n$ operators. By using eq. (\ref{deltaSigma}) we get 
 \be 
 \nu=\frac{3}{2} \sqrt{1- 8 \, c_L^2 \frac{(1+w)}{(1+3 \, w)^2}}\, ;
 \ee
the isentropic part of the power spectrum can be derived again from our general results  (\ref{power_spectrum}) and (\ref{specrel})
\be
\begin{split}
&{\cal P}_{{\cal R}^{(0)}}=\frac{2^{2\, \nu-4}\, c_L^{2-2\,\nu} \,	\Gamma(\nu)^2 \, H_{in}^{-\frac{4}{(1+3 \,w)}}}{\epsilon_i \, \pi^3 \, M_{pl}^2 } (-t)^{1-2\, \nu-\frac{4}{(1+3 \,w)}} \; k^{3-3\, \sqrt{1- 8 \, c_L^2 \,\frac{(1+w)}{(1+3 \, w)^2}}} \, , \\
&{\cal P}_{{\zeta}^{(0)}}= \frac{(1+3\,w)^2}{4\, c_L^4} \,{\cal P}_{{\cal R}^{(0)}}\, .
\end{split}
\ee
In order to reproduce the measured~\cite{Komatsu:2008hk,Akrami:2018odb} red spectral tilt $n_s< 1$, a negative $c_L^2$ is needed as shown in  figure (\ref{ns}).
Such result is in agreement with~\cite{Sitwell:2013kza}.
Thus, $w$-Media inflation for a zero temperature solid is ruled out being the spectrum
inevitably blue tilted when $c_L^2>0$.

Consider now the effect of $\delta \sigma$. In the case of a $w$-Media,
$c_b^2=w$ and the condition (\ref{BDC}) is trivially satisfied;
the time dependence of the source term in (\ref{caneq2}) is
\be
\gamma=-\frac{6 \,w}{(1+3\,w)}\, , \qquad \Sigma_k= (-1)^{\gamma}\,
\frac{\varphi'_{in}}{\sqrt{1+ w}}\,\left[\frac{(1+3 \,w)\,(c_L^2-w)}{4\, c_L}\right] \,\delta \hat \sigma\, . 
\ee 
The size of $\Sigma_k$ is basically set by $\delta \sigma$ unless $w
\approx -1$. From  $(\ref{spectrum-deltasigma})$ we get
\be
{\cal P}^{(\delta \sigma)}\propto \frac{|\Sigma_k|^2\; k^3}{\epsilon}\, H_{in}^{-\frac{4}{1+3\,w}} \, ;
\ee
moreover
\be
\frac{{\cal P}^{(\delta \sigma)}}{{\cal P}_{\zeta^{(0)}}} \sim \frac{|\delta \sigma|^2 }{\epsilon_i}\,k^{2\, \nu}\, (-t)^{-1+2\,\nu+\frac{4}{(1+3 \,w)}}\, .
\ee
Such a ratio is time dependent if we take an almost flat tilt for ${\cal
  P}_{\zeta^{(0)}}$ with $\nu \approx \frac{3}{2}$
\be
\frac{{\cal P}^{(\delta \sigma)}}{{\cal P}_{\zeta^{(0)}}} \sim \frac{|\delta \sigma|^2 }{\epsilon_i}\,k^{2\, \nu} \,(-t)^{6\frac{(1+w)}{1+3\,w}}\, .
\ee
The conclusion is
that the entropic contribution is in general dominant;  thus a
blue-tilted spectrum can turned in a red-tilted one by taking
$\delta \sigma$ such that ${\cal P}^{(\delta\sigma)}$ is the dominant
part of the spectrum with a red tilt. We will study such
possibility in a separate paper~\cite{future} where the dynamics of $\delta \sigma$ will be analyzed in detail.

\section{Gravitational Waves}

\label{G_W}
Spin two perturbations are defined by
\be
g_{00} =- a^2 \, , \qquad g_{0i} =0 \, , \qquad g_{ij} = a^2
\left(\delta_{ij} + \chi_{ij}\right) \, .
\ee
For all media considered the dynamics of spin two modes during
inflation can be derived from the following Lagrangian (Fourier basis)
  \be
  L_T = \frac{a^2 \, \plm^2}{2} \left[ \chi_{ij}' \chi_{ij}' -
    \left(k^2 + \frac{M_2}{a^2} \right) \chi_{ij} \chi_{ij}  \right] \, .
  \ee
The mass parameter $M_2$ in $c_T^2$, which in the scalar sector controls the amount of the anisotropic
  stress, it is also behind the
dispersion relation for the propagation of the gravitational waves. By 
using (\ref{infpar}), the equation of motion  can be written  as
\be
  \chi_{ij}''+\frac{2 \,\beta}{t}\chi_{ij}'+\chi_{ij}\left[k^2
    +\frac{6 \, c_T^2 \, \beta^2 \, (1+w)}{t^2}\right]=0 \, .
\ee
As usual, the quantum field $\chi_{ij}$ is  made up of two polarizations 
\be 
\chi_{ij}= \sum_{s=1}^2 \varepsilon_{ij}^s\, \chi_s \, , \qquad \chi_s=\chi \, \hat{a}^s_k+\chi^* \,\hat a^{s\,\dagger}_k \, ,
\ee
with modes
\be
\begin{split}
&\chi={\cal C}_1\, t^{\frac{1}{2} (1-2\beta)}
H_{\nu_T}^{(1)}(-k \,t)+{\cal C}_2 \, t^{\frac{1}{2} (1-2\,\beta)}
H_{\nu_T}^{(2)}(-k \,t), \\
& \nu_T=\frac{1}{2} \sqrt{1-4\,\beta(1-\beta)-24 \, c_T^2 \, \beta^2 \, (1+w)}\, .
\end{split}
\ee
Taking into account a factor two due to polarization average, we get
for the primordial tensor power spectrum 
\be
{\cal P}_{T}=\frac{k^3}{\pi^2} |\chi|^2=\frac{4^{\nu_T-1}\, \Gamma(\nu_T)^2\, H_{in}^{-2\,\beta}}{\pi^3 \, M_{pl}^2}( -t)^{1-2\,\beta -2\,\nu_T} \, k^{3-2\,\nu_T}\, . 
\ee
Once the tensor power spectrum is given, by using our previous results,
we can compute the tensor to scalar ratio in the case of ${\cal R}$
and $\zeta$ separately.
 The entropic contribution is not in general suppressed by $ \epsilon$ as for 
single field slow-roll inflation, but determined by
$\delta\sigma$.
The ratio is defined as
\be
r_A= \frac{{\cal P}_T}{{\cal P}_A}=\frac{{\cal P}_T}{{\cal P}_A^{(0)}+{\cal P}_A^{(\delta\sigma)}} \, .
\ee
The suffix $A$ stands for $A=\zeta, \; {\cal R} $; indeed, being the
Weinberg theorem typically violated the two power spectra must be
studied separately. Taking  $\delta\sigma$ of order $O(\sqrt{\epsilon_i})$, the
two contributions of the total power spectrum are in general comparable
with some notable exception. 
Hereafter, we will use the following definitions/relations
\be
\frac{1}{r_{A}}=\frac{1}{r_{A}^{(0)}}+ \frac{1}{r^{(\delta\sigma)}}\, , \qquad r_A^{(0)}= \frac{{\cal P}_T}{{\cal P}_A^{(0)}}\,, \qquad r^{(\delta\sigma)}= \frac{{\cal P}_T}{{\cal P}^{(\delta\sigma)}}\,.
\ee
Obviously there are two extreme regimes
\begin{enumerate}
\item $r_A^{(0)} \ll r^{(\delta\sigma)} \, \Rightarrow \, \frac{{\cal P}^{(\delta\sigma)}}{{\cal P}_A^{(0)}} \ll 1$ and $r_{A}\approx r_{A}^{(0)}$ ,
\item $r_A^{(0)} \gg r^{(\delta\sigma)}\,\Rightarrow \, \frac{{\cal P}^{(\delta\sigma)}}{{\cal P}_A^{(0)}} \gg 1$ and $r_{A}\approx r^{(\delta \sigma)}$ .
\end{enumerate}
When the isentropic part is sub-dominant, the ratio is more or less
unchanged with respect to the single field slow-roll scenario. If the
entropic part is the dominant one, the ratio will be even more
suppressed. Thus, the more interesting case is when the two
contributions are comparable and this happens respectively when
\begin{itemize}
\item SSR: $c_b^2=\frac{2}{3}$;
\item SR: $c_b^2=-1$;
\item WM: If we consider $\nu \approx \frac{3}{2}$, the entropic part always dominates unless $w$ is not close to minus one.
\end{itemize}
Details are given in appendix \ref{GW_ratio}.

\section{Post-Inflation Evolution: Instantaneous Reheating}

\label{Post-Inf}
 The violation of the Weinberg theorem makes the inflationary regime less
 robust; after inflation ends, ${\cal R}$ and $\zeta$ will be not conserved for
superhorizon modes, thus the reheating phase becomes crucial to make
quantitative predictions on how the initial conditions for the hot
universe era are set. More than analyzing specific reheating models we will try to
 focus as much as possible on general features, approximating reheating as 
 instantaneous and taking place on a suitable time-like hypersurface where the Israel junction
conditions~\cite{Israel:1966rt} will be imposed.
The transition hypersurface ${\cal T}$ is given in terms of a 4-dimensional
scalar $q$ as $q=$constant, or expanding at the linear order in
perturbation theory
\be
\bar q+\delta q= \text{constant} \, . 
\ee
We will denote by the subscript $f_-$ the  quantity $f$ evaluated at the end of
inflation, while with $f_+$ the same quantity evaluated at the end of
reheating when the medium that had driven inflation has decayed into the
standard components of the hot Universe. The change of $f$
across ${\cal T}$ will be  written as $[f]_{\cal T}= f_+-f_-$.
On  ${\cal T}$,  the equation
of state suddenly changes from the value at end of inflation $w_-=-1+2
\, \epsilon_f/3$ to $w_+=1/3$ in the radiation dominated phase. 
The junction conditions~\cite{Deruelle:1995kd} can be written as 
\begin{enumerate}
\item \textbf{First Junction condition}:\\[.2cm]
 $[h_{a b}]_{\cal T}=0$\\[.2cm]
  The induced metric $h_{ab}$ on ${\cal T}$ is continuous. In
  particular, at the background level this is equivalent to
  the continuity of the scale factor $a$. At the linear level
  the jump of $\phi$ is determined by $\delta q$ according with 
\be
 \left[-\phi+{\cal H} \frac{\delta q}{\bar q'}\right]_{\cal T}=\left[\zeta_{q}\right]_{\cal T}=0 \, .
\ee
Namely,
\be
\zeta_q=-\phi+{\cal H} \frac{\delta q}{\bar q'}
\label{zetaqdef}
\ee
represents the gauge invariant curvature perturbation
of a constant $q$ hypersurfaces  and it is continuous across ${\cal T}$. 
\item \textbf{Second Junction condition}\\[.2cm]
$\left[K_{ab}\right]_{\cal T}=0$\\[.2cm]
We have assumed that no surface layer EMT tensor is present. In the absence of a localized EMT on ${\cal T}$, the extrinsic curvature $K_{ab}$ has to be continuous. At the background level, it implies the continuity of ${\cal H}$ across ${\cal T}$, namely $[{\cal
    H}]=0$. At the linear level from its diagonal part
\be
\left[\left(1-\frac{{\cal H}'}{{\cal H}^2}\right) \left({\cal R}- \zeta_{q}\right) \right]_{\cal T}=0 \, ,
\label{zetaq}
\ee
while from the off-diagonal part
\be
\left[\frac{\delta q}{\bar{q}'}\right]_{\cal T}=0\, .
\ee
Together with the continuity of $\zeta_q$ leads to the continuity of $\phi$.
Thus, the perturbation $\delta q$ can be characterized by a sharp jump during the transition. 
\end{enumerate}
The choice of no surface layer EMT is consistent with the form of our
EMT: though $M_2$ and $\delta\sigma$ can have a finite jump, no
$\delta$-like contribution is present.
From the continuity of $\zeta_q$, the junction condition (\ref{zetaq})
tells that ${\cal R}$ must jump across ${\cal T}$~\footnote{Being
  $\zeta_q$ continuous on ${\cal T}$ we have that
  $\zeta_{q\,-}=\zeta_{q\,+} \equiv \zeta_q$.} 
\be
\label{R_after_re}
\epsilon_+\,\left({\cal R}_{+}-\zeta_q \right)= \epsilon_{f} \left({\cal
    R}_{-}-\zeta_q\right)\qquad \Rightarrow \qquad {\cal
  R}_{+}=\zeta_q +\frac{\epsilon_f}{\epsilon_+} \left({\cal
    R}_{-}-\zeta_q\right) \, ,
\ee
A crucial point is that when the Weinberg theorem holds,
whatever the hypersurface, $\zeta_{q}$ is conserved at superhorizon scales and $\zeta_{q} \approx \zeta_{q'}|_{q\neq q'}$; therefore the
choice of the curvature perturbation does not matter.  
However, if the Weinberg theorem is violated, the quantity
continuous across ${\cal T}$ will depend on the choice of $q$. In other words, the
violation of the Weinberg theorem inevitably introduces a certain
degree of dependence on the details of reheating.

If, as it usually done,  one assumes that the instantaneous reheating
takes place at $\rho=$const.~\cite{Endlich:2012pz} an issue emerges. From one hand,  a
medium-driven inflation will produce ${\cal R} \neq \zeta$  and 
  at $\rho=$const. will transmit such a
difference. On the other hand, a
$\Lambda$CDM universe is incompatible with ${\cal R} \neq
\zeta$. Indeed, the Weinberg theorem holds and the only source of
non-conservation for ${\cal R}$ and  $\zeta$ comes from entropic
perturbations encoded in $\Gamma_{\Lambda\text{CDM}}$ defined by the
following superhorizon dynamics
\be
{\cal R}'=\zeta' = \frac{{\cal H} \, \Gamma_{\Lambda\text{CDM}}}{\rho
  +p} \, ,  \qquad \text{with  } {\cal R} = \zeta\, .
\ee
In general $\Gamma_{\Lambda\text{CDM}}$ is different from zero
when the various species have different density perturbations in the
presence of isocurvature modes; see for instance~\cite{Celoria:2017xos}.
Let us discuss the isentropic ($\delta \sigma =0$) and the general
adiabatic case separately.

\subsection{Isentropic Inflation}

Let us first assume $\delta\sigma=0$ and $\rho$=const. reheating.
For the inflationary regimes of interest we have the following  situation.
\begin{enumerate}
\item \textbf{Super slow-roll}\\
The value of $\epsilon_f$ is so small that ${\cal
    R}_{+}=\zeta$; the comoving curvature sharply jumps assuming as
  final value the one of $\zeta$ at end of inflation.
 Differently from~\cite{Chen:2013kta}, the distinction between $\zeta$
 and ${\cal R}$ is crucial. 
 This is the only regime  where reheating taking place at  constant energy surface
 does not transmit the difference between ${\cal R}$ and $\zeta$ to the radiation dominated epoch.

\item \textbf{Slow-roll}\\
 At the end of inflation ${\cal R}$ and $\zeta$ are different if
 $c_L^2 \neq 1$, see eq. (\ref{zetaise}), and 
 \be
\label{R_solid}
{\cal R}_{+}=\zeta+ \frac{\epsilon_f}{\epsilon_+}\left({\cal R}_- 
-\zeta\right)=
\zeta \left[ 1- \frac{\epsilon_f}{2}(c_L^2+1)\right] \, .
\ee
As a result, a small difference between ${\cal R}$ and $\zeta$ is transmitted to the radiation phase suppressed by $\epsilon_f$. This difference cannot be
 explained using relative entropy perturbations 
 after reheating in the standard $\Lambda$CDM.
Such a  difference can be attributed only to a particular dark
sector~\cite{Celoria:2017xos,Celoria:2017idi} where  the
Weinberg theorem is violated with or without intrinsic entropy perturbations.
\item \textbf{$w$-Media}\\
 $\epsilon_f$ is not in general small and the violation of the
 Weinberg theorem is still sizable and the same considerations for
 slow-roll applies.
\end{enumerate} 
\subsection{Adiabatic Case}
Let us now discuss the general case whit $\delta \sigma \neq 0$.
The choice of a constant energy density hypersurface is dynamically
justified only when entropy perturbations are absent during
 the two cosmological epochs. However, when the Weinberg
theorem is violated this is not the only choice.
The idea is to 
identify the scalar $q$  that  can be considered constant during  the  transition. 
Suppose that the current $J^\mu= n \, u^\mu$ is conserved during reheating, intuitively we can imagine that the total relative number density $\frac{\delta n}{n}$ is continuous during the instantaneous transition, namely
\be
\left[\frac{\delta n}{\bar{n}}\right]=0\, .
 \label{piL_con2}
\ee
A dynamical derivation of the  continuity  of $\delta n/ \bar{n}$ is given in appendix \ref{zeta_n}.
Taking as $n=$constant the reheating hypersurface, the junction
conditions imply that the scalar curvature 
\be
\label{conN_curv}
 \zeta_n=-\phi +{\cal H}\frac{\delta n}{\bar n'}\, ,
\ee
is continuous during the transition.
After some manipulations, one can show that $\zeta_n$ is related to
longitudinal phonons during inflation, namely
\be
\zeta_n- = \frac{k^2 }{3} \pi_L= \zeta-  
\frac{\bar\varphi'\,\delta\sigma}{3\, a^4\, \rho\,(1+w)} \, .
\ee
The very same quantity, at the end of the reheating, when the universe
is radiation dominated, is given by  
\be
\zeta_n =\zeta -\frac{\Gamma_{\Lambda\text{CDM}}}{3\, \rho\,(1+w)\,
  c_s^2}
\, ;
\ee
where $\Gamma_{\Lambda\text{CDM}}$ is the total entropic contribution
coming from the possible  difference in the density perturbations of the
various components of the universe when isocurvature modes are present~\footnote{In
  $\Lambda$CDM all components are taken as barotropic fluids and there are non intrinsic entropic perturbation.}.
Once again,  if the transition takes place between two adiabatic phases
for which $\Gamma=0$, $\zeta_n=\zeta$ is continuous and the
difference between ${\cal R}$ and $\zeta$ is transmitted to the
radiation dominated phase.\\
In the transition at $n=$const. the ``standard'' $\zeta$ is not
continuous and its jump is fixed by the condition $[\zeta_n]=0$, while from
(\ref{zetaq}) we can find  the final value  ${\cal R}_+$ of
the comoving curvature perturbation at end of reheating. Thus 
\be
\begin{split}
\label{jumps}
&[\zeta]=\left[\frac{ \Gamma_{\text{eff}}}{3 \,\rho\, (1+w)\, c_s^2}\right]\, ,\\
&{\cal R}_+= \zeta_n+\frac{\epsilon_f}{\epsilon_+}\left({\cal R}_--\zeta_n\right)\, .
\end{split}
\ee
The effective quantity $\Gamma_{\text{eff}}$ is defined as follows:
\be
\text{Inflation: }\Gamma_{\text{eff}}=-\frac{\bar\varphi'}{a^4}\, c_s^2 \, \delta\sigma\,, \qquad \text{Radiation: } \Gamma_{\text{eff}}=\Gamma_{\Lambda\text{CDM}}\, .
\ee 
Imagine that during inflation the intrinsic entropy perturbation per particle $\delta \sigma$ is exactly zero. In this case
\be
\zeta_n=\zeta_{n\, -}=\zeta^{(0)}_- \, , \qquad \Gamma_{\text{eff}}{}_+=\Gamma_{\Lambda\text{CDM}}\, . 
\ee
As a consequence  (\ref{jumps}) reduces to
\be
\begin{split}
&\zeta_+=\zeta^{(0)}_-+\frac{ \Gamma_{\Lambda\text{CDM}}}{3 \,\rho\, (1+w)\, c_s^2}|_+\\
&{\cal R}_+= \zeta^{(0)}_-+\frac{\epsilon_f}{\epsilon_+}\left({\cal R}^{(0)}-\zeta^{(0)}\right)|_-\, .
\end{split}
\ee
Now we take into account that, in the radiation phase for $\Lambda$CDM, we can show that\cite{Celoria:2017xos} at superhorizon scales: 
\be
{\cal R}_+-\zeta_+=-\frac{\Gamma_{\Lambda\text{CDM}}}{3\, \rho\, (1+w)\, c_s^2}|_++\frac{{\cal R}'}{3\, {\cal H}\, c_s^2}|_+=0 \, . 
\ee 
Consequently, we get the following crucial result
\be
\label{res}
{\cal R}_+=\zeta_+=
\zeta^{(0)}_-+\frac{\epsilon_f}{\epsilon_+}\left({\cal
    R}^{(0)}-\zeta^{(0)}\right)|_-\, ,
\ee
with
the value of $\Gamma_{\Lambda\text{CDM}}$  fixed by inflation
dynamics according with
\be
\frac{\epsilon_f}{\epsilon_+} \left(\zeta^{(0)}-{\cal R}^{(0)}\right)|_-+\frac{\Gamma_{\Lambda\text{CDM}}}{3\, \rho \, (1+w)\,c_s^2}|_+ =0\, .
\label{Gamma1}
\ee
The superhorizon value of ${\cal R}$  and
$\Gamma_{\Lambda\text{CDM}}$ given by given by (\ref{res}) and
(\ref{Gamma1}) can be used to set the initial conditions for the standard
evolution of the universe.
As anticipated, the bottom line is that the violation of the Weinberg
theorem inflation  triggers  the generation of
relative entropy perturbations, namely isocurvature modes, in $\Lambda$CDM.

Finally, let us consider the effect of an
intrinsic entropy perturbations during inflation. In this case we have
that 
\be
-\frac{\Gamma_{\text{eff}}}{3\, \rho \, (1+w)\,c_s^2}|_- = \alpha\, \zeta^{(\delta\sigma)}|_-\, , \qquad \zeta_n=\zeta_-^{(0)}+(1+\alpha)\zeta_-^{(\delta\sigma)}\,.
\ee
The value of the constant $\alpha$ is generically of order
  one, and its precise value depends on $c_b^2$. In this case,
(\ref{jumps}) gives 
\be
\begin{split}
&\zeta_+=\zeta^{(0)}_-+(1+\alpha)\,\zeta^{(\delta\sigma)}_-+\frac{\Gamma_{\Lambda\text{CDM}}}{3\, \rho \, (1+w)\,c_s^2}|_+\;\\
&{\cal R}_+= \zeta^{(0)}_-+(1+\alpha)\,\zeta^{(\delta\sigma)}_- +\frac{\epsilon_f}{\epsilon_+}\left({\cal R}^{(0)}-\zeta^{(0)}-\alpha\, \zeta^{(\delta\sigma)}\right)|_-\, .
\end{split}
\ee
If we consider $\Lambda$CDM,  we get a result similar to (\ref{res}),
with an additional
contribution due to the presence of $\delta \sigma$
\be
\label{res2}
\begin{split}
&\zeta_+={\cal R}_+=\zeta^{(0)}_-+(1+\alpha)\,\zeta^{(\delta\sigma)}_- +\frac{\epsilon_f}{\epsilon_+}\left({\cal R}^{(0)}-\zeta^{(0)}-\alpha\, \zeta^{(\delta\sigma)}\right)|_-\, ;\\
&  \frac{\epsilon_f}{\epsilon_+}\left(\zeta^{(0)}+\alpha\,
  \zeta^{(\delta\sigma)}-{\cal R}^{(0)}\right)|_-
+\frac{\Gamma_{\Lambda\text{CDM}}}{3\, \rho \, (1+w)\,c_s^2}|_+=0\, ;
\end{split} 
\ee
where we have used that ${\cal R}^{(\delta\sigma)}=\zeta^{(\delta\sigma)}$.
Note that also the intrinsic $\zeta^{(\delta \sigma)}_-$ part is
always suppressed by $\epsilon_f$.
If we consider that after reheating with a very good precision accuracy: $\epsilon_+ \approx 2\,, \; w_+\approx 1/3\, , \; c_s^2{}_+\approx 1/3$, we get that the $\Gamma_{\Lambda\text{CDM}}$ is  
\be 
\label{fin_res}
\Gamma_{\Lambda\text{CDM}}= \rho\left(1+w\right) |_-\, \left({\cal R}-\zeta-\alpha\,\zeta^{(\delta\sigma)}\right)|_{-}\, .
\ee
In conclusion, in the case of  inflation driven by an isentropic
medium, the choice of surface where the instantaneous reheating takes
place is crucial to determine how the difference $({\cal R}-\zeta)$
is transmitted to the radiation phase. For both  the choices of constant $\rho$ or
$n$ hypersurface for reheating, the corresponding  curvature
perturbation $\zeta$ or $\zeta_n$ is continuous or characterized by a
small jump proportional to $\epsilon_f$. On the contrary,  the
comoving curvature perturbation ${\cal R}$ sharply jumps in trying to
match the value of $\zeta$ or $\zeta_n$ respectively.
These above mentioned feature has a dramatic consequence on what scalar one uses to
define scale invariant power spectrum: the only reasonable choice is $\zeta$\footnote{In the general adiabatic case $\zeta_n$ will be the correct choice, however being at this level $\zeta_n^{(\delta\sigma)} \propto \zeta^{(\delta\sigma)}$ and $\zeta_n^{(0)}=\zeta^{(0)}$, the additive entropic contribution to the power spectrum has the same identical features of what we have called ${\cal P}^{(\delta \sigma)}$ in the previous sections.}
from which the tilt and tensor to scalar ratio are extracted. 
\section{Conclusions}
\label{Concl}
In this paper,  by using an
effective field theory approach, we have studied in a systematic way
inflation  driven by a generic
adiabatic medium, extending the previous works on fluid and solid
inflation. The focus is on the violation of the Weinberg theorem
during inflation which has two main sources: anisotropic
stress (for solids) and entropic perturbations. A third mechanism,
related to the peculiar background dynamics of the super slow-roll regime, plays a major role for
fluids for which the standard slow-roll regime is not
viable.   In general, even if a
media is adiabatic, a constant in time entropy per particle $\sigma(\vec{x})$ can
be  present. As a result, when the medium is
non-barotropic, the entropy per particle perturbation $\delta \sigma$ acts
as a source term in the dynamical equation for both the comoving curvature
perturbation ${\cal R}$ and the uniform curvature perturbation $\zeta$.
At superhorizon scales, both ${\cal R}$ and  $\zeta$ are {\it not} constant in time
with  $\zeta \neq {\cal R}$. At the level of power spectrum,
we get that the scalar amplitude of momentum $k$ keeps growing until inflation ends,
even if $k$ is superhorizon, moreover an  additional contribution
due to  $\delta \sigma$ is present. Being
$\delta \sigma$ strictly constant for an adiabatic medium, its
contribution to the power spectrum is an undetermined function of the
comoving momentum. It is interesting to note that, when the adiabaticity
condition is removed, $\delta \sigma$ will be dynamically determined
with  $\delta \sigma'=0$ at superhorizon scales; thus,   an adiabatic
medium can be considered as a convenient simplified model
to describe a generic non-dissipative medium  at superhorizon
scales.  
The violation of the Weinberg theorem is also relevant for the
reheating phase. Considering  an instantaneous reheating,
taking place on a suitable spacelike hypersurface, the variation
of $\zeta$ and ${\cal R}$ is essentially geometric and encoded in the
Israel junction conditions.

In the case $\delta \sigma=0$ (  both during inflation and radiation domination ) energy conservation
suggests that the instantaneous reheating takes place on $\rho =$
constant  hypersurface  with    $\zeta$  continuous 
 while ${\cal R}$ jumps.   In fluid
inflation, the power spectrum of $\zeta$ is almost scale free
but it requires  $c_s^2 \gg 1$,
which makes fluid inflation not very exciting. 
In the slow-roll case,
a small violation of the Weinberg theorem during the
radiation phase is present, resulting in a difference $({\cal
  R}-\zeta)|_+$ which is  order $\epsilon$. Similar
considerations  apply to the case of  $w$-media inflation,
with the only difference that $\epsilon$ is not small in general,
however it predicts a blue-tilted power spectrum which 
 is  ruled out by observations. 

  The violation of the Weinberg theorem is also important during
reheating. We have shown that, in the instantaneous approximation, reheating on
$\rho=$constant hypersurface makes strictly speaking impossible to
connect  inflation with a standard radiation dominated $\Lambda$CDM universe. Indeed, to take
into account the difference of $\zeta$ and ${\cal R}$ a
non-standard dark energy sector is needed. Insisting on $\Lambda$CDM
requires to reconsider the reheating hypersurface. Our analysis shows that
in the contest of adiabatic media is rather natural to take the
constant particle number hypersurface across which the corresponding
curvature perturbation $\zeta_n$ is continuous .  
Applying the junction conditions we arrive at the conclusion that the
difference $({\cal R}-\zeta)|_-$ during inflation  and any intrinsic entropic contribution $\zeta^{(\delta\sigma)}$ induces
a contribution of isocurvature perturbations on the top of the
adiabatic ones for $\Lambda$CDM radiation dominated epoch.
Furthermore, in slow roll, the change of $\zeta$ is small and order $\epsilon$.
Consequently, we can still use the $\zeta$ power spectrum to define
the tilt parameter and the tensor to scalar ratio.
It is also interesting to point out that the additive entropic contribution
to the power spectrum can be used as seed to form primordial black
holes that could constitute an important part of dark matter in the universe. 
Finally, let us briefly comment on primordial non-Gaussianity. 
The local $f_{NL}$ parameter in the squeezed configuration which is
most constrained  by observations~\cite{Akrami:2019izv}, plays a
crucial role in the CMB angular
bispectrum~\cite{Bartolo:2011wb,Pajer:2013ana,Cabass:2016cgp}, and in
the scale dependence of the halo bias~\cite{Bartolo_2005,PhysRevD.77.123514,dePutter:2015vga}. Here,
the validity of the Weinberg Theorem is strictly related with the
validity of the Maldacena consistency relations for the squeezed
configuration in single field inflation~\cite{Maldacena:2002vr}. Thus the violation of the Weinberg theorem directly implies the possibility to get features which are considerably different from the single field inflation predictions.
While primordial non-Gaussianity for an isentropic
solid was already computed in~
\cite{Endlich:2012pz}, the case for a general media which has two
scalar modes  is presently unknown and we hope to report on it soon.

\newpage

\appendix
 \section{Masses}
\label{massesapp}

The explicit values of the mass parameters in terms of $U$ and its
derivatives, are given in the case of a FLRW spacetime in conformal
time by
\be
\label{MXY}
\begin{split}
& M_0= \frac{\phi'^2}{2 \, \plm^2}
 \left[a^2 \left(U_{Y^2}-2\;
   U_X\right)-4 \;a\; \phi '\; U_{XY}+4 \;U_{X^2} \, \phi'^2\right] 
, \\
& M_1= \frac{2\, \phi'^2}{\plm^2} \left(\sum_{n=0}^3 \;  a^{-2 \;n}\;
   U_{y_n}+a^2\;  U_{X}\right),\\
& M_2= - \frac{2}{\plm^2}\; \sum_{n=1}^3\;  n^2\,a^{2
   (2-n)} \;U_{\tau _n},\\
& M_3=\frac{1}{\plm^2} \left( 
     2\sum_{m,n=1}^3 m\, n\,
   a^{-2 m-2 n+4}\, U_{\tau _m \tau
   _n}+2 \sum_{n=1}^3 n\, a^{1-2 n}\;
   \,U_{b \tau _n}-
    \sum_{n=1}^3 
    a^{4-2 n}\, U_{\tau _n}+
   \frac{1}{2\,a^2} U_{b^2}  \right) \, ,\\
&  M_4= \frac{\phi'}{2\,\plm^2} \left(2\,  \sum_{n=1}^3 
   a^{3-2 n}\,U_{Y \tau
   _n}-a^3\, U_Y+   U_{bY}\right) \\
&  \,\; \qquad +
  \frac{\phi'^2}{\plm^2} \left(-2\,  \sum_{n=1}^3\,  a^{2-2 n}
   \;U_{X \tau _n}+ a^2\, U_{X}- \frac{1}{a}\,
   U_{b X}\right) ;
\end{split}
\ee
%

\section{\texorpdfstring{$\zeta$}{Lg} Evolution}
\label{app:zeta}
Proceeding as for ${\cal R}$ one can write down an evolution equation for
$\zeta$ only which has the form
\be
\left[ \zeta' \, \frac{6 a^2 (w+1) \mathcal{H}^2}{2 k^2+9 (w+1)
    \mathcal{H}^2} \right]' + m_\zeta^2 \, \zeta + B_\sigma \, \delta
\hat \sigma =0 \, ,
\label{zetaeq}
\ee
with
\be
\begin{split}
  & m_\zeta^2 = 6 \, a^2 \,  \mathcal{H}^2 \, (1+w) \frac{ \left[ 2  \, k^4 \,
    c_L^2+9 \, k^2\,  \mathcal{H}^2 \,  (w+1)
  \left(8 \, c_T^2-1\right)+54  \, \mathcal{H}^4 \, (w+1)^2
   \, c_T^2 \right] }{\left[2 \, k^2+9 \, (w+1) \,
   \mathcal{H}^2\right]^2} \, ;\\
&  B_\sigma = \bar{\varphi}' \frac{4 \, k^4 (c_L^2- c_b^2) +4 \,  k^2
  \, 
  \, F_2 + 9 \, (1+ w) \, F_0}{ \left[6 \left(2 k^2+9 (w+1) \mathcal{H}^2\right)^2  \right]} \\
& F_2  = \mathcal{H}^2 \left[ 3 \, c_b^2 \left(3
        \, c_b^2-3 \, c_L^2+4 \, c_T^2-2\right)+6 \, c_L^2 +12 \, w \, c_T^2+4 \, c_T^2 \right ]+\mathcal{H} \left(3 \,
   c_L^2{}'-4 \, 
   c_T^2{}'-3 \, c_b^2{}'\right) \, ; \\
 & F_0 =  {\cal H}^4 \Big \{ c_b^2 \left(48 \, c_T^2-36 \, c_L^2+9\,
   w-9\right)+18 \, 
   c_b^4+3\,  w \left(8 \, c_T^2-3 \, c_L^2\right)+2 \left(3 \, 
     c_L^2-4 \, c_T^2\right){}^2+9 \, c_L^2 \Big \} \\
   & \quad  + {\cal H}^3 \left[6( c_L^2{}' - \, c_b^2{}')-8 \, c_T^2{}' \right] \, .
\end{split}
\ee

\section{Dynamics of entropy perturbations on superhorizon scales}
\label{sigma_time_dep}
According to~\cite{Celoria:2017bbh},  when
$M_1\neq 0$, at superhorizon scales the entropy per particle perturbation $\delta\sigma$ evolution is described by
\be
\left[\delta \sigma' \frac{\bar \varphi'{}^2 \left(6  \,a^2 \, (1+w \,){\cal H}^2+M_1\right)}{6\, a^2 \, M_1 (1+w)\, {\cal H}^2}\right]'=0 \,, \qquad \frac{k^2}{{\cal H}^2} \ll 1 \, ,
\ee
which can be easily integrated to give 
\be
\delta\sigma=\delta\sigma_0+ \delta\sigma_1\, \int^t dt' \, \frac{1}{M_{pl}\, \varphi'{}^2}\, \frac{6\, a^2 \, (1+w)\, {\cal H}^2\, M_1}{M_1+6\, a^2 \, (1+w)\, {\cal H}^2}\, .
\ee
Let us parametrize the $M_1$ time dependence as follows
\be
M_1=m_1\, a^\lambda_1 \, . 
\ee
If we assume the constant $m_1$ to be not too small, we get at the leading order in the slow-roll parameter $\epsilon_i$
\be
\delta\sigma=\delta\sigma_0+\delta\sigma_1 \frac{4 \, \epsilon_i \, H_{in}^{-6\, c_b^2 (1+\epsilon_i)-\eta}}{\left[1+6\, c_b^2 (1+\epsilon_i)+\eta\right]\, M_{pl} \, \varphi_{in}'{}^2}\,(-t)^{-1-6\, c_b^2 (1+\epsilon_i)-\eta} . 
\ee
The above  result holds in the SSR and SR regimes. Thus, the absence of entropy growing modes is ensured by the condition\footnote{We do not analyze WM regime, where the result will be naturally $M_1$ dependent.} 
\be
c_b^2 < -\frac{(1+\eta)}{6(1+\epsilon_i)} \, \Rightarrow\; \text{SSR}\,:\; c_b^2 <-\frac{1}{6}(1+\eta), \;\;\text{SR}\,:\; c_b^2<-\frac{1}{6}\,. 
\ee
Note that this relation is compatible with (\ref{BDC}).  

\section{Tensor to scalar ratio}
\label{GW_ratio}
Let us analyze the tensor to scalar ratio $r$ in the relevant
inflationary regimes.
For fluid inflation in the super slow-roll regime, when $\beta=-1$ and
$w \approx -1$, we get $\nu_T=3/2$ and the tensor spectrum is exactly scale-free with $n_T=1$
\be 
{\cal P}_T=\frac{1}{2}\frac{H_{in}^2}{\pi^2\, M_{pl}^2} \, .
\ee
For  $r$, depending on the value of $c_b^2$, we have the following possible cases. 
\begin{enumerate}
\item $\frac{2}{3} < c_b^2 < 3$\\
For both $\zeta$ and ${\cal R}$, we have ${\cal P}^{(\delta \sigma)} \ll
  {\cal P}^{(0)}$ and thus the total scalar is dominated by the isentropic part $r^{(0)}$. In particular
\be
\begin{split}
& r_{\zeta}^{(0)}= 9\, 2^{10+\eta} \frac{\Gamma\left(\frac{3}{2}\right)^2\, c_s^{-5-\eta}}{\Gamma\left(-\frac{(5+\eta)}{2}\right)^2}\left(-k\, t\right)^{-(10+\eta)} \, \epsilon(t) \, ,\\
& r_{\cal R}^{(0)}= 2^{7+\eta}\frac{\Gamma\left(\frac{3}{2}\right)^2\, c_s^{-5-\eta}}{\Gamma\left(-\frac{(3+\eta)}{2}\right)^2}\left(-k\, t\right)^{-(6+\eta)} \, \epsilon(t)\,.
\end{split}
\ee
The difference between ${\cal R}$ and $\zeta$ is
important, indeed $r_{\cal R}^{(0)}/r_{\zeta}^{(0)}\sim (-k\,t)^4$; thus 
 $r_{\cal R}^{(0)}$  is much smaller then $r_{\zeta}^{(0)}$  in the
superhorizon limit. This was not noticed in~\cite{Chen:2013kta}.
\item $c_b^2=\frac{2}{3}$\\
${\cal P^{\delta \sigma}} \sim {\cal P}^{(0)}$. The entropic part
is comparable with the standard one. 
\item $c_b^2 < \frac{2}{3}$\\
In this case, ${\cal P}^{(\delta \sigma)} \gg {\cal P}^{(0)}$
and thus the total $r$ is close to $r^{(\delta\sigma)}$ with
\be 
r^{(\delta\sigma)} \sim  \,
\frac{\epsilon_i^2}{|\delta\sigma|^2} \,k^{-3}\,t^{-6 \,(1+c_b^2)-2 \,\eta}\, ,
\ee
setting $\eta=-10$ for a flat scalar spectrum, again for $|k\,t|\ll 1$ the
ration goes to zero very fast.
\end{enumerate}
Clearly, even in the presence of entropic perturbations, in fluid
inflation the tensor to scalar ratio is vanishingly small.\\
Consider now slow-roll inflation for solids. In this case 
\be
\nu_T=\frac{3}{2}+\epsilon-\frac{4}{3}  \,\epsilon \, c_T^2 \, ,  \qquad {\cal P}_T= \frac{1}{2} \,\frac{H_{in}^2}{\pi^2\, M_{pl}^2} \; t^{\frac{8}{3} \,\epsilon  \,c_T^2} \; k^{-2 \,\epsilon+\frac{8}{3}  \,\epsilon \, c_T^2} . 
\ee
Again, $c_b^2$ is an important parameter for the tensor to scalar ratio.
\begin{enumerate}
\item $-1 < c_b^2 < -\frac{1}{3}$\\
The isentropic part of the  power spectrum is dominant and $r$ will be
practically equal to $r^{(0)}$ with
\be
r^{(0)}_{\cal R}=4\, c_L\, \epsilon_i \,k^{\eta}\, , \qquad r^{(0)}_{\zeta}=r_{\cal R}\,c_L^4  .
\ee 
\item  $c_b^2=-1$\\
${\cal P^{\delta \sigma}} \sim {\cal P}^{(0)}$. The entropic part
is comparable with the standard one. 
\item $c_b^2 < -1$\\
 The entropic part of the power spectrum is dominant and of the order
 \be
r^{(\delta\sigma)} \sim  \frac{\epsilon_i^2}{|\delta \sigma|^2 }  \,k^{-3} \,
(-t)^{-6 \,(1+c_b^2)}  \, ;
\ee
it is completely analogous to  fluid inflation and  the tensor to
scalar ratio is strongly suppressed.
\end{enumerate}
Finally, let us consider $w$-media inflation, for which 
\be
\nu_T=-\frac{3}{2 \,(1+3 \, w)} \,\sqrt{(w-1)^2-8 \,(1+w) \,(c_L^2-w)}=\nu  \, .
\ee
Taking for instance $\nu_T \approx \frac{3}{2}$ which gives a blue
tilt with $c_L^2 >0$
\be
{\cal P_T}=\frac{H_{in}^{-2 \,\beta}\, (-t)^{-\frac{1}{2}-2 \,\beta}}{2\, \pi^2\, M_{pl}^2} \, . 
\ee
Being $\nu_T=\nu$, the isentropic tensor to scalar ratio is time and scale independent
\be
r^{(0)}_{\cal R}= \frac{6 \,(w+1)}{c_L^{2 \,(1+\nu)}} \, , \qquad r_{\zeta}= r^{(0)}_{\cal R}\, \frac{4  \,c_L^4}{(1+ 3 \, w)^2} \, . 
\ee
As discussed in the previous sections, being in this case ${\cal
  P}^{(\delta \sigma)} \gg {\cal P}^{(0)}$, the entropic
contribution will be the dominant one and the computation of
$r^{(\delta\sigma)}$ will be necessary
\be
r^{(\delta\sigma)} \sim \frac{1}{(1+3 \, w)^2} \frac{\epsilon_i^2}{|\delta\sigma|^2} \, k^{-3} \, (-t)^{-6 \, \frac{(1+w)}{1+3 \,w}}\, .
\ee

\section{Reheating hypersurface}
\label{zeta_n}

As   discussed in section \ref{Post-Inf}, the choice of a
constant energy density as the reheating hypersurface is not the only
option when the Weinberg theorem is violated.
Following~\cite{Mukhanov:991646}, we will study  how reheating can take place on a different surface. 
Let us start, by using the $00$ component of the linearized
Einstein equations, we get 
\be
{\cal R}-\zeta= \frac{2}{9}\frac{k^2 \, \phi}{ (1+w) \,{\cal H}^2} \, ,
\label{Rzetarel1}
\ee
We can generalize eq. (\ref{Rzetarel}) as follows
\be
\label{Rprime}
{\cal R}'= 3  \,{\cal H}\, c_L^2 \, \left({\cal R}  \, \,\frac{c_s^2}{c_L^2}-\zeta\right)+\frac{ {\cal H}\,\tilde\Gamma}{\rho \,(1+w)} \, . 
\ee
In the hydrodynamical approximation, the above relation holds
  before and after inflation. During inflation scalar anisotropic
  stress is encoded in $c_L^2$ through $c_T^2$ and 
\be
\tilde\Gamma=\frac{\varphi'}{a^4} \,(c_b^2-c_L^2) \, \delta \sigma\, .
\ee
During radiation domination, taking  for simplicity $\Lambda$CDM
constituted by a collection of isentropic fluids, anisotropic terms
are absent and $c_L \to c_s$, while the $\tilde\Gamma$ reduces to
\be
\tilde\Gamma=\Gamma_{\Lambda\text{CDM}}=\Gamma_{rel}\, . 
\ee 
Integrating (\ref{Rprime}) along the transition, we can estimate the
jump of ${\cal R}$, extending the analysis of~\cite{Mukhanov:991646};
namely 
\be
 \left[{\cal R}\right]= \int_{-}^{+} c_s^2 \, \theta^2  \,\left(\frac{2}{3}  \,\frac{a^2}{{\cal H}} \,k^2\,\phi+\frac{ {\cal H}\,\Gamma_{\text{eff}}}{\rho\,(1+w) \, c_s^2 \,\theta^2 }\right) \, dt \, ,
\ee
where $\theta$ is given by
\be
\theta=\frac{1}{a\,\sqrt{1+w}}\, . 
\ee
$\Gamma_{\text{eff}}$ stands for the part of the $\Gamma$ which can give a contribution during the integration, i.e:
\be
\text{Inflation: }\Gamma_{\text{eff}}=-\frac{\bar\varphi'}{a^4}\, c_s^2 \, \delta\sigma\,, \qquad \text{Radiation: } \Gamma_{\text{eff}}=\Gamma_{\Lambda\text{CDM}}\, .
\ee 
Using the relation
\be 
c_s^2\, \theta^2=\frac{\rho}{3\, a^2\, {\cal H}} \left(\frac{1}{\rho+p}\right)'-\frac{\rho}{a^2(\rho+p)}\, ,
\ee
and integrating by parts we get
\be
\label{Zeta_cont}
\left[{\cal R}\right]=\left[\frac{1}{3}  \,\frac{k^2  \,\phi}{{\cal H}^2 \,\epsilon}+\frac{ \Gamma_{\text{eff}}}{3 \,\rho\, (1+w)\, c_s^2}\right]\, , 
\ee
which, thanks to (\ref{Rzetarel1}), is equivalent to
\be
\label{piL_con}
\left[\zeta-\frac{ \Gamma_{\text{eff}}}{3 \,\rho\, (1+w)\, c_s^2}\right]=[\zeta_q]=0\, . 
\ee
Eq. (\ref{piL_con}) can also be interpreted  the continuity of
the curvature of  a suitable $q=$const. hypersurface during the transition.
By definition, we get
\be
\frac{\delta q}{\bar q'}= \frac{\delta \rho}{\bar
  \rho'}-\frac{ \Gamma_{\text{eff}}}{3 \,\rho\,{\cal H}\, (1+w)\, c_s^2}\, .
\ee
As a  matter of fact, $q$ is a linear combination of $\rho$ and
total entropy per particle $\sigma$. In particular during inflation we
can take $q=n$, where $n$ is the medium number density; thus 
\be
\zeta_n=\frac{k^2}{3} \pi_L=\zeta+ \frac{\bar\varphi'\, \delta\sigma}{12\, a^2\, \epsilon\,{\cal H}^2\, M_{pl}^2} =-\phi+{\cal H} \, \frac{\delta n}{\bar n'} \, .
\ee
On the other hand, during radiation domination, one can show that
taking $n=\sum_i \,n_i (\rho_i,\sigma_i)$, one gets 
\be
\zeta_n{}_+= \zeta_+-\frac{ \Gamma_{\Lambda\text{CDM}}}{3 \,\rho\, (1+w)\, c_s^2}\, ;
\ee
the sum in the definition of $n$ refers to the different fluid
components during $\Lambda$CDM.
Thus, we have just demonstrated that $\zeta_n$ is continuous in the
transition from inflation to radiation domination~\footnote{Note that eq. (\ref{piL_con}) is
  completely general and holds also in the non-adiabatic case when
  entropy perturbations can propagate.}, coherently with the
assumption in section \ref{Post-Inf}
\be
\left[\frac{\delta n}{n}\right]=-3\,\left[{\cal H}\,\frac{\delta n}{\bar n'}\right] =0\, .
\ee

\section*{Acknowledgement}
We thank Sabino Matarrese for very useful discussions.

\bibliographystyle{unsrt}  
\bibliography{infl}

\begin{thebibliography}{10}

\bibitem{Weinberg:2003sw}
S.~Weinberg.
\newblock {Adiabatic modes in cosmology}.
\newblock {\em Phys. Rev.}, D67:123504, 2003.

\bibitem{Weinberg:2008zzc}
S.~Weinberg.
\newblock {Cosmology}.
\newblock {\em Oxford Univ. Press}, 2008.

\bibitem{Cheung:2007st}
Clifford Cheung, Paolo Creminelli, A.~Liam Fitzpatrick, Jared Kaplan, and
  Leonardo Senatore.
\newblock {The Effective Field Theory of Inflation}.
\newblock {\em JHEP}, 03:014, 2008.

\bibitem{Weinberg:2008hq}
Steven Weinberg.
\newblock {Effective Field Theory for Inflation}.
\newblock {\em Phys. Rev.}, D77:123541, 2008.

\bibitem{Kinney:2005vj}
William~H. Kinney.
\newblock {Horizon crossing and inflation with large eta}.
\newblock {\em Phys. Rev.}, D72:023515, 2005.

\bibitem{Namjoo:2012aa}
Mohammad~Hossein Namjoo, Hassan Firouzjahi, and Misao Sasaki.
\newblock {Violation of non-Gaussianity consistency relation in a single field
  inflationary model}.
\newblock {\em EPL}, 101(3):39001, 2013.

\bibitem{Motohashi:2014ppa}
Hayato Motohashi, Alexei~A. Starobinsky, and Jun'ichi Yokoyama.
\newblock {Inflation with a constant rate of roll}.
\newblock {\em JCAP}, 1509(09):018, 2015.

\bibitem{Akhshik:2015nfa}
M.~Akhshik, H.~Firouzjahi, and S.~Jazayeri.
\newblock {Effective Field Theory of non-Attractor Inflation}.
\newblock {\em JCAP}, 1507(07):048, 2015.

\bibitem{Celoria:2017xos}
Marco Celoria, Denis Comelli, and Luigi Pilo.
\newblock {Intrinsic Entropy Perturbations from the Dark Sector}.
\newblock {\em JCAP}, 1803(03):027, 2018.

\bibitem{Chen:2013kta}
X.~Chen, H.~Firouzjahi, M.~H. Namjoo, and M.~Sasaki.
\newblock {Fluid Inflation}.
\newblock {\em JCAP}, 1309:012, 2013.

\bibitem{Endlich:2012pz}
S.~Endlich, A.~Nicolis, and J.~Wang.
\newblock {Solid Inflation}.
\newblock {\em JCAP}, 1310:011, 2013.

\bibitem{Matarrese:1984zw}
S.~Matarrese.
\newblock {On the Classical and Quantum Irrotational Motions of a Relativistic
  Perfect Fluid. 1. Classical Theory}.
\newblock {\em Proc. Roy. Soc. Lond.}, A401:53--66, 1985.

\bibitem{Dubovsky:2005xd}
S.~Dubovsky, T.~Gregoire, A.~Nicolis, and R.~Rattazzi.
\newblock {Null energy condition and superluminal propagation}.
\newblock {\em JHEP}, 03:025, 2006.

\bibitem{Dubovsky:2011sj}
S.~Dubovsky, L.~Hui, A.~Nicolis, and D.T. Son.
\newblock {Effective field theory for hydrodynamics: thermodynamics, and the
  derivative expansion}.
\newblock {\em Phys. Rev.}, D85:085029, 2012.

\bibitem{Ballesteros:2012kv}
G.~Ballesteros and B.~Bellazzini.
\newblock {Effective perfect fluids in cosmology}.
\newblock {\em JCAP}, 1304:001, 2013.

\bibitem{Ballesteros:2016kdx}
G.~Ballesteros, D.~Comelli, and L.~Pilo.
\newblock {Thermodynamics of perfect fluids from scalar field theory}.
\newblock {\em Phys. Rev.}, D94(2):025034, 2016.

\bibitem{Celoria:2017bbh}
Marco Celoria, Denis Comelli, and Luigi Pilo.
\newblock {Fluids, Superfluids and Supersolids: Dynamics and Cosmology of Self
  Gravitating Media}.
\newblock {\em JCAP}, 1709(09):036, 2017.

\bibitem{Bird:2016dcv}
Simeon Bird, Ilias Cholis, Julian~B. Muñoz, Yacine Ali-Haïmoud, Marc
  Kamionkowski, Ely~D. Kovetz, Alvise Raccanelli, and Adam~G. Riess.
\newblock {Did LIGO detect dark matter?}
\newblock {\em Phys. Rev. Lett.}, 116(20):201301, 2016.

\bibitem{Sasaki:2016jop}
Misao Sasaki, Teruaki Suyama, Takahiro Tanaka, and Shuichiro Yokoyama.
\newblock {Primordial Black Hole Scenario for the Gravitational-Wave Event
  GW150914}.
\newblock {\em Phys. Rev. Lett.}, 117(6):061101, 2016.
\newblock [erratum: Phys. Rev. Lett.121,no.5,059901(2018)].

\bibitem{Clesse:2016ajp}
Sebastien Clesse and Juan García-Bellido.
\newblock {Detecting the gravitational wave background from primordial black
  hole dark matter}.
\newblock {\em Phys. Dark Univ.}, 18:105--114, 2017.

\bibitem{Espinosa:2017sgp}
J.~R. Espinosa, D.~Racco, and A.~Riotto.
\newblock {Cosmological Signature of the Standard Model Higgs Vacuum
  Instability: Primordial Black Holes as Dark Matter}.
\newblock {\em Phys. Rev. Lett.}, 120(12):121301, 2018.

\bibitem{future}
Marco Celoria, Denis Comelli, Luigi Pilo, and Rocco Rollo.
\newblock {In Preparation}.
\newblock 2019.

\bibitem{ussgf}
G.~Ballesteros, D.~Comelli, and L.~Pilo.
\newblock {Massive and modified gravity as self-gravitating media}.
\newblock {\em Phys. Rev.}, D94(12):124023, 2016.

\bibitem{Celoria:2017hfd}
Marco Celoria, Denis Comelli, and Luigi Pilo.
\newblock {Sixth mode in massive gravity}.
\newblock {\em Phys. Rev.}, D98(6):064016, 2018.

\bibitem{Guth:1980zm}
Alan~H. Guth.
\newblock {The Inflationary Universe: A Possible Solution to the Horizon and
  Flatness Problems}.
\newblock {\em Phys. Rev.}, D23:347--356, 1981.
\newblock [Adv. Ser. Astrophys. Cosmol.3,139(1987)].

\bibitem{Linde:1981mu}
Andrei~D. Linde.
\newblock {A New Inflationary Universe Scenario: A Possible Solution of the
  Horizon, Flatness, Homogeneity, Isotropy and Primordial Monopole Problems}.
\newblock {\em Phys. Lett.}, 108B:389--393, 1982.
\newblock [Adv. Ser. Astrophys. Cosmol.3,149(1987)].

\bibitem{Albrecht:1982wi}
Andreas Albrecht and Paul~J. Steinhardt.
\newblock {Cosmology for Grand Unified Theories with Radiatively Induced
  Symmetry Breaking}.
\newblock {\em Phys. Rev. Lett.}, 48:1220--1223, 1982.
\newblock [Adv. Ser. Astrophys. Cosmol.3,158(1987)].

\bibitem{Langlois:2008qf}
David Langlois, Sebastien Renaux-Petel, Daniele~A. Steer, and Takahiro Tanaka.
\newblock {Primordial perturbations and non-Gaussianities in DBI and general
  multi-field inflation}.
\newblock {\em Phys. Rev.}, D78:063523, 2008.

\bibitem{Arroja:2008yy}
Frederico Arroja, Shuntaro Mizuno, and Kazuya Koyama.
\newblock {Non-gaussianity from the bispectrum in general multiple field
  inflation}.
\newblock {\em JCAP}, 0808:015, 2008.

\bibitem{Celoria:2017idi}
Marco Celoria, Denis Comelli, and Luigi Pilo.
\newblock {Self-gravitating $\Lambda$-media}.
\newblock {\em JCAP}, 1901(01):057, 2019.

\bibitem{Ballesteros:2018wlw}
Guillermo Ballesteros, Jose Beltran~Jimenez, and Mauro Pieroni.
\newblock {Black hole formation from a general quadratic action for
  inflationary primordial fluctuations}.
\newblock {\em JCAP}, 1906(06):016, 2019.

\bibitem{Komatsu:2008hk}
E.~Komatsu et~al.
\newblock {Five-Year Wilkinson Microwave Anisotropy Probe (WMAP) Observations:
  Cosmological Interpretation}.
\newblock {\em Astrophys. J. Suppl.}, 180:330--376, 2009.

\bibitem{Akrami:2018odb}
Y.~Akrami et~al.
\newblock {Planck 2018 results. X. Constraints on inflation}.
\newblock {\em Submitted to A\&A}, 2018.

\bibitem{Sitwell:2013kza}
Michael Sitwell and Kris Sigurdson.
\newblock {Quantization of Perturbations in an Inflating Elastic Solid}.
\newblock {\em Phys. Rev.}, D89(12):123509, 2014.

\bibitem{Israel:1966rt}
W.~Israel.
\newblock {Singular hypersurfaces and thin shells in general relativity}.
\newblock {\em Nuovo Cim.}, B44S10:1, 1966.
\newblock [Nuovo Cim.B44,1(1966)].

\bibitem{Deruelle:1995kd}
Nathalie Deruelle and Viatcheslav~F. Mukhanov.
\newblock {On matching conditions for cosmological perturbations}.
\newblock {\em Phys. Rev.}, D52:5549--5555, 1995.

\bibitem{Akrami:2019izv}
Y.~Akrami et~al.
\newblock {Planck 2018 results. IX. Constraints on primordial non-Gaussianity}.
\newblock 2019.

\bibitem{Bartolo:2011wb}
N.~Bartolo, S.~Matarrese, and A.~Riotto.
\newblock {Non-Gaussianity in the Cosmic Microwave Background Anisotropies at
  Recombination in the Squeezed limit}.
\newblock {\em JCAP}, 1202:017, 2012.

\bibitem{Pajer:2013ana}
Enrico Pajer, Fabian Schmidt, and Matias Zaldarriaga.
\newblock {The Observed Squeezed Limit of Cosmological Three-Point Functions}.
\newblock {\em Phys. Rev.}, D88(8):083502, 2013.

\bibitem{Cabass:2016cgp}
Giovanni Cabass, Enrico Pajer, and Fabian Schmidt.
\newblock {How Gaussian can our Universe be?}
\newblock {\em JCAP}, 1701(01):003, 2017.

\bibitem{Bartolo_2005}
Nicola Bartolo, Sabino Matarrese, and Antonio Riotto.
\newblock Signatures of primordial non-gaussianity in the large-scale structure
  of the universe.
\newblock {\em Journal of Cosmology and Astroparticle Physics},
  2005(10):010--010, oct 2005.

\bibitem{PhysRevD.77.123514}
Neal Dalal, Olivier Dor\'e, Dragan Huterer, and Alexander Shirokov.
\newblock Imprints of primordial non-gaussianities on large-scale structure:
  Scale-dependent bias and abundance of virialized objects.
\newblock {\em Phys. Rev. D}, 77:123514, Jun 2008.

\bibitem{dePutter:2015vga}
Roland de~Putter, Olivier Doré, and Daniel Green.
\newblock {Is There Scale-Dependent Bias in Single-Field Inflation?}
\newblock {\em JCAP}, 1510(10):024, 2015.

\bibitem{Maldacena:2002vr}
Juan~Martin Maldacena.
\newblock {Non-Gaussian features of primordial fluctuations in single field
  inflationary models}.
\newblock {\em JHEP}, 05:013, 2003.

\bibitem{Mukhanov:991646}
Viatcheslav Mukhanov.
\newblock {\em {Physical Foundations of Cosmology}}.
\newblock Cambridge Univ. Press, Cambridge, 2005.

\end{thebibliography}

\end{document}